\documentclass[
aps,
superscriptaddress,
nofootinbib,
longbibliography,
preprint,
amsmath,
amssymb,
10pt
]{revtex4-2}
\usepackage[utf8]{inputenc}

\usepackage{graphicx}  
\usepackage{dcolumn}   

\usepackage[dvipsnames]{xcolor} 
\usepackage[colorlinks = true,
            linkcolor  = RoyalBlue,
            urlcolor   = RoyalBlue,
            citecolor  = RoyalBlue]{hyperref}

\usepackage{amsmath} 
\usepackage{amssymb} 
\usepackage{bbm}     

\usepackage{lineno} 
\usepackage{xspace} 

\newcommand{\XeXe}         {\mbox{Xe--Xe}\xspace}
\newcommand{\PbPb}         {\mbox{Pb--Pb}\xspace}

\newcommand{\pt}           {\ensuremath{p_{\rm T}}\xspace}

\newcommand{\twosevensixnn}{$\sqrt{s_{\mathrm{NN}}}=2.76$~TeV\xspace}
\newcommand{\fivenn}       {$\sqrt{s_{\mathrm{NN}}}=5.02$~TeV\xspace}
\newcommand{\fivefourfournn}{$\sqrt{s_{\mathrm{NN}}}=5.44$~TeV\xspace}

\let\gevc=\GeVc
\let\mevc=\MeVc
\let\tev=\TeV

\let\mev=\MeV

\newcommand{\fluidum}      {\textsc{Fluid{\it u}M}\xspace}

\newcommand{\trento}       {\textsc{TrENTo}\xspace}
\newcommand{\fastreso}     {\textsc{FastReso}\xspace}
\begin{document}

\title{Mapping properties of the quark gluon plasma in Pb-Pb and Xe-Xe collisions at energies available at the CERN Large Hadron Collider}
\date{\today}

\author{L.~Vermunt}
\email[]{luuk.vermunt@cern.ch}
\affiliation{Physikalisches Institut, Universit{\"a}t Heidelberg, 69120 Heidelberg, Germany}
\affiliation{GSI Helmholtzzentrum f{\"u}r Schwerionenforschung, 64291 Darmstadt, Germany}

\author{Y.~Seemann}
\email[]{yannis.seemann@gmail.com}
\affiliation{Physikalisches Institut, Universit{\"a}t Heidelberg, 69120 Heidelberg, Germany}

\author{A.~Dubla}
\email[]{a.dubla@cern.ch}
\affiliation{GSI Helmholtzzentrum f{\"u}r Schwerionenforschung, 64291 Darmstadt, Germany}

\author{S.~Floerchinger}
\email[]{stefan.floerchinger@uni-jena.de}
\affiliation{Theoretisch-Physikalisches Institut
Friedrich-Schiller-Universität Jena, 07743 Jena, Germany} 

\author{E.~Grossi}
\email[]{eduardo.grossi@unifi.it}
\affiliation{Dipartimento di Fisica, Universit\`a di Firenze and INFN Sezione di Firenze,
50019 Sesto Fiorentino, Italy}

\author{A.~Kirchner}
\email[]{kirchner@thphys.uni-heidelberg.de }
\affiliation{Institut f\"ur Theoretische Physik Heidelberg, 69120 Heidelberg, Germany} 

\author{S.~Masciocchi}
\email[]{s.masciocchi@gsi.de}
\affiliation{Physikalisches Institut, Universit{\"a}t Heidelberg, 69120 Heidelberg, Germany}
\affiliation{GSI Helmholtzzentrum f{\"u}r Schwerionenforschung, 64291 Darmstadt, Germany}

\author{I.~Selyuzhenkov} 
\email[]{ilya.selyuzhenkov@gmail.com}
\affiliation{GSI Helmholtzzentrum f{\"u}r Schwerionenforschung, 64291 Darmstadt, Germany}

\begin{abstract}
A phenomenological analysis of the experimental measurements of transverse momentum spectra of identified charged hadrons and strange hyperons in \PbPb and \XeXe collisions at the LHC is presented.
The analysis is based on the relativistic fluid dynamics description implemented in the numerically efficient \fluidum approach.
Building on our previous work, we separate in our treatment the chemical and kinetic freeze-out, and incorporate the partial chemical equilibrium to describe the late stages of the collision evolution.
This analysis makes use of Bayesian inference to determine key parameters of the QGP evolution and its properties including the shear and bulk viscosity to entropy ratios, the initialisation time, the initial entropy density, and the freeze-out temperatures.
The physics parameters and their posterior probabilities are extracted using a global search in multidimensional space with modern machine learning tools, such as ensembles of neural networks.
We employ our newly developed fast framework to assess systematic uncertainties in the extracted model parameters by systematically varying key components of our analysis.
\end{abstract}

\maketitle
\tableofcontents

\newpage  

\section{Introduction}

Heavy-ion collisions at ultra-relativistic energies have been at the forefront of modern physics research for several decades. 
These collisions, studied at facilities such as the Relativistic Heavy Ion Collider (RHIC) and the Large Hadron Collider (LHC), create an extreme state of matter known as the quark--gluon plasma (QGP)~\cite{Busza:2018rrf, ALICE:2010suc, STAR:2005gfr, PHENIX:2004vcz}.
This fluid is of great interest because it is described by a renormalisable and fundamental quantum field theory at the microscopic level, i.e.~quantum chromodynamics (QCD). 
While the macroscopic fluid properties remain challenging to calculate from first principles and a limited number of computations exist to date, 
an increasing number of experimental results motivate phenomenological and theoretical studies.
One of the key features of the QGP is its collective behaviour, which is thought to arise from the fluid-like properties of the system.
However, recent experimental observations of collective behaviour in proton--nucleus and proton--proton collision systems~\cite{ALICE:2019zfl, PHENIX:2018lia, CMS:2015fgy, ATLAS:2015hzw} have challenged the uniqueness of the fluid-like response of the QGP. 
The resolution of the origins of collective behaviour in heavy-ion collisions will likely rely on a quantitative rather than just qualitative agreement between data and models.

Traditionally, physical properties of QCD matter have been determined by comparing experimental data with model calculations of event-averaged and predefined observables. 
However, recent developments have shown promise in extracting more information from the final-state particles produced in heavy-ion collisions. 
Two such methods are Bayesian analysis~\cite{Moreland:2018gsh, Bernhard:2019bmu, JETSCAPE:2020shq, JETSCAPE:2020mzn, Nijs:2020ors, Nijs:2020roc, Nijs:2021clz, Nijs:2023yab, Heffernan:2023utr, Liyanage:2023nds, Parkkila:2021tqq, Parkkila:2021yha} and deep learning~\cite{Pang:2016vdc, Steinheimer:2019iso}. 
Bayesian analysis uses global fitting to simultaneously determine multiple model parameters, utilising all available experimental data. 
On the other hand, deep learning identifies observables that are sensitive to specific physical properties, enabling the extraction of relevant information. 
This work improves upon our previous analysis of Ref.~\cite{Devetak:2019lsk} by using Bayesian inference as the optimality criterion. 
The previous approach of minimising the $\chi^2$-value had several limitations, including the neglect of experimental data correlations, interpolation uncertainties, and parameter correlations. 
In addition, we improved our theoretical description with respect to Ref.~\cite{Devetak:2019lsk} by separating the chemical and kinematic freeze-out, and incorporating the partial chemical equilibrium to describe the later stages of the evolution, as well as by exploiting a parametrisation from Yang--Mills theory for the shear viscosity to entropy ratio of the QGP~\cite{Christiansen:2014ypa,Pawlowski_private} instead of using a constant value.

Our model for simulating high-energy nuclear collisions combines three distinct components. 
The \trento model~\cite{Moreland:2014oya} was utilised for the initial conditions, while the \fluidum model~\cite{Floerchinger:2018pje}, featuring a mode splitting technique for very fast computations, was used for the relativistic fluid dynamic expansion with viscosity. 
Additionally, the \fastreso code~\cite{Mazeliauskas:2018irt} was used to take resonance decays into account. 
Despite \fluidum being significantly faster than event-by-event hydrodynamic codes, it is still beneficial to develop an emulator that can function as a quick substitute for the full model. 
Where previous Bayesian analyses~\cite{Moreland:2018gsh, Bernhard:2019bmu, JETSCAPE:2020shq, JETSCAPE:2020mzn, Nijs:2020ors, Nijs:2020roc, Nijs:2021clz, Nijs:2023yab, Heffernan:2023utr, Liyanage:2023nds} utilised Gaussian process regression, we for the first time exploit neural network emulation. 
It has been demonstrated in Ref.~\cite{Neal1996} that infinitely wide neural networks can approximate Gaussian process regression. 
Furthermore, the use of neural networks has several computational benefits, including significantly reduced training time, lower memory usage, and the ability to handle any number of inputs and outputs. This newly developed framework will have additional applications in the precise fitting of charm and beauty observables contributing to further constrain the heavy-quark spatial diffusion coefficient, widening our knowledge on QGP transport properties~\cite{Capellino:2023cxe,Capellino:2022nvf}. 

In the present work, we determine the specific shear and bulk viscosity to entropy ratios of the QGP, as well as the freeze-out temperatures $T_{\rm kin}$ and $T_{\rm chem}$, the starting time of a fluid description $\tau_0$, and the normalisation of the initial entropy profile. 
We employ our newly developed fast framework to comprehensively explore systematic uncertainties in the extracted model parameters by systematically varying critical components of our analysis. 
We compare against experimental measurements of transverse momentum spectra of identified charged hadrons ($\pi$, $\rm K$, $\rm p$) and strange hyperons ($\Lambda$) with $\pt < 2$~\gevc in \PbPb collisions at \twosevensixnn and \fivenn, and \XeXe collisions at \fivefourfournn from the ALICE Collaboration~\cite{ALICE:2013mez,ALICE:2013cdo,ALICE:2019hno,ALICE:2021lsv}.
Given that this paper primarily serves as a proof of concept for our new Bayesian inference framework, our study will be restricted to the 0--5\% centrality range.
We specifically chose a very central bin because we expect the background--fluctuation splitting ansatz that underlies \fluidum to work best for central collisions, where the profiles tend to be the most symmetric.
Furthermore, the restriction to only one centrality class was imposed such that we are able to keep the initial-state parameters, which play an important role in the centrality dependence of the observables, fixed.
In a forthcoming publication, we will explore the use of anisotropic flow observables, as well as experimental data from different centrality classes, to gain further insight into the key parameters of the QGP.
Nevertheless, it is important to underscore that through the analysis of transverse momentum spectra exclusively, we can derive significant constraints on essential aspects such as the bulk viscosity, the freeze-out temperatures, $\tau_0$, and the normalisation.
This work highlights the potential of such analyses to provide valuable insights into the properties of the QGP.

This paper is organised as follows.
We summarise the details of the initial conditions, the hydrodynamic evolution, and the hadronisation procedures in Sec.~\ref{sec:setup}.
The procedure of the Bayesian analysis is discussed in Sec.~\ref{sec:analysis}.
We then discuss the extraction of the model parameters in Sec.~\ref{sec:results}, and end with a summary and future directions in Sec.~\ref{sec:summary}.

\section{Modelling of the different stages of a heavy-ion collision}
\label{sec:setup}

The following section provides a brief overview of the various components of our theoretical model. 
We first examine the initial conditions, which are determined using the \trento model~\cite{Moreland:2014oya}. 
Subsequently, we turn to the time evolution as implemented in \fluidum~\cite{Floerchinger:2018pje}, which solves the equations of relativistic fluid dynamics with shear and bulk viscosity and corresponding relaxation times. 
The newly introduced partial chemical equilibrium will be discussed next, together with the kinetic freeze-out and the implementation of strong resonance decays performed with \fastreso~\cite{Mazeliauskas:2018irt}.

\subsection{Initial conditions: \trento}
\label{sec:initialcond}

As in our previous work~\cite{Devetak:2019lsk}, we use the \trento model parametrisation~\cite{Moreland:2014oya} for the initial conditions. 
This is an effective model, intended to generate realistic Monte Carlo initial transverse entropy (or energy) profiles without assuming specific physical mechanisms. 
It involves positioning nucleons with a Gaussian width $w$ using a fluctuating Glauber model, while ensuring a minimum distance $d$ between them. 
Each nucleon contains $m$ randomly placed constituents with a Gaussian width of $v$. 
\trento uses an entropy deposition parameter $p$ that interpolates among qualitatively different physical mechanisms for entropy production~\cite{Moreland:2014oya}. 
Furthermore, additional multiplicity fluctuations are introduced by multiplying the density of each nucleon by random weights sampled from a gamma distribution with unit mean and shape parameter $k$.

For this study, the \trento parameters are not estimated via the Bayesian analysis. 
Instead, they are set based on the current state of knowledge in literature. 
As reviewed extensively in Ref.~\cite{Giacalone:2022hnz}, we set $w=0.5$~fm, $m=4$, $v=0.4$~fm, $p=0$, and use the \trento output as entropy density. 
In addition, we set $k=1$ and $d=0.75$~fm, based on the outcome of the Bayesian analysis of Ref.~\cite{Nijs:2020ors}. 
For the nucleon--nucleon cross section, the measurements by the ALICE Collaboration~\cite{ALICE-PUBLIC-2018-011} are used, i.e.~$x=61.8, \, 67.6,$ and $68.4$~mb for \PbPb at \twosevensixnn, \PbPb at \fivenn, and \XeXe at \fivefourfournn, respectively. 
The Pb ion is sampled from a spherically symmetric Woods--Saxon distribution with radius $R=6.65$~fm and surface thickness $a=0.54$~fm, while the Xe ion comes from a deformed spheroidal Woods--Saxon distribution with $R=5.60$~fm, $a=0.49$~fm, and deformation parameters $\beta_2=0.21$ and $\beta_4=0.0$~\cite{Bally:2021qys}. 
Using this set of parameters (which we call the ``central'' configuration in the remaining) we have generated the transverse density $T_{\rm R}(x,y)$ for $1.5\cdot10^6$ minimum-bias events with impact parameter sampled from the range $b \in [0 \, {\rm fm} , 20 \, {\rm fm}]$.

To classify the \trento events in centrality classes, we utilise the integrated transverse density $\int {\rm d}^2x T_{\rm R}(\vec x)$, which is expected to have a linear monotonic relationship with multiplicity~\cite{Giacalone:2020ymy}. 
This allows us to divide the events into narrow multiplicity classes of one percent, each of which can be treated as an ensemble of events with random orientation in the transverse plane as in experiment.
For each centrality class, we determine the averaged or expected entropy density profile as 
\begin{equation}
    s(r)=\frac{{\rm Norm}}{\tau_0} \left\langle T_{\rm R}(r) \right\rangle,
    \label{eq:sinitial}
\end{equation}
where the average $\langle \cdots \rangle$ is taken over all the events in the class with a random reaction plane angle.
Consequently, the average is independent of the azimuthal angle $\phi$ by construction.
We have introduced a centrality class-dependent normalisation constant ${\rm Norm}$ to account for possible variation of the fits from the \trento multiplicity scaling. 
We account for the longitudinal expansion effect (Bjorken flow) at early times by scaling the ${\rm Norm}$ by the initialisation time $\tau_0$.
To enlarge the statistic of the average entropy profile $\langle T_{\rm R}(r)\rangle$ (as we will see in the next section, it can be taken independent of $\phi$ for this work), we perform the event average over all events in one centrality class
\begin{equation}
    \langle T_{\rm R}(r)\rangle = \frac{1}{2\pi} \int_0^{2\pi} \mathrm{d}\phi \;  \left\langle T_{\rm R}(r,\phi) \right\rangle
\end{equation}
As in Ref.~\cite{Devetak:2019lsk}, we produce the averaged entropy densities for the larger experimental centrality classes by averaging the corresponding distributions from the more narrow classes and propagating those.

Note that a free-streaming phase that evolves the energy density profile at $\tau = 0$ for a short time scale to finite $\tau_0$, is currently not included in our framework.
To mimic the effect of a free-streaming phase, which can be seen as a suppression/dilution of the ``spikiness'' of the initial entropy density profiles, we also test another set of \trento parameters (called the ``freesteaming'' configuration later on) in which the width of the nucleons is increased (i.e.~$w=1.0$~fm, $v=0.9$~fm, and $d=1.25$~fm). 
A larger nucleon width in \trento is observed to suppress the spikiness in the initial entropy density.
Systematic variations of the other \trento parameters are not considered.

\subsection{Hydrodynamic evolution: \fluidum}
\label{sec:fluidum}

The software package \fluidum~\cite{Floerchinger:2018pje}, which utilises a theoretical framework based on relativistic fluid dynamics with mode expansion~\cite{Floerchinger:2013rya, Floerchinger:2013hza, Floerchinger:2014fta}, is used to solve the equations of motion for relativistic fluids. 
This involves decomposing the fluid fields into background and fluctuation components. 
Specifically, the fluid fields are represented as $\Phi(\tau, r, \phi, \eta) = \Phi_{0}(\tau, r) + \epsilon \, \Phi_{1}(\tau, r, \phi, \eta)$, where $\Phi_{0}$ is the background solution, $\Phi_{1}$ is the perturbation around it, and $\epsilon$ is the formal expansion parameter (set to $\epsilon \to 1$ at the end). 
The background and perturbations can be solved accurately and efficiently using numerical algorithms~\cite{Floerchinger:2018pje}.
The background--fluctuation splitting is justified due to two statistical symmetries that are approximately observed in high-energy nuclear collisions.
Firstly, the collision energy is sufficiently high, leading to an approximate boost invariance of the system at midrapidity.
This implies that observables at midrapidity exhibit only a mild dependence on the rapidity, and consequently, the dependence from it can be a perturbation. 
Secondly, there is a statistical symmetry related to the random orientation of the reaction plane angle.
In each collision event, the two ions collide with a particular relative orientation in the reaction plane, which is uncorrelated with other events. 
When calculating observables averaged over multiple events, the azimuthal angle's dependence is effectively removed, unless the reaction plane angle is explicitly reconstructed and corrected for each event.
Moreover, the average will always be independent of the azimuthal angle, and every dependence form is a perturbation.

For our study, we are interested in examining the azimuthally averaged transverse momentum spectra of identified particles at midrapidity. 
Therefore, we do not consider azimuthally and rapidity-dependent perturbations and only require the background solution to the fluid evolution equations, neglecting terms of quadratic or higher order in perturbation amplitudes.
The causal equations of motion are obtained from second-order Israel--Stewart hydrodynamics~\cite{Floerchinger:2017cii}.

\begin{figure*}[tb!]
    \includegraphics[width=0.478\linewidth]{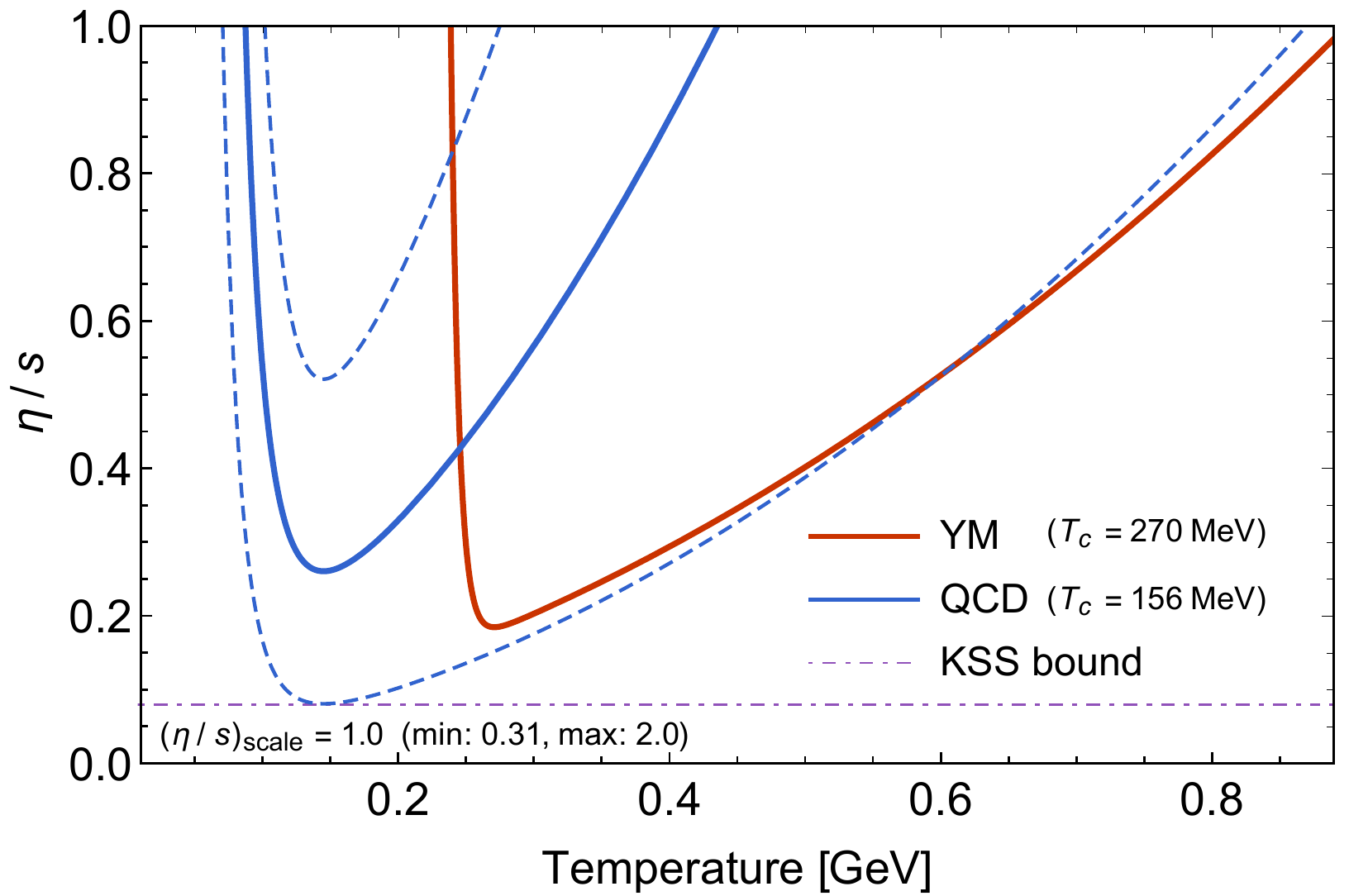}
    \includegraphics[width=0.5\linewidth]{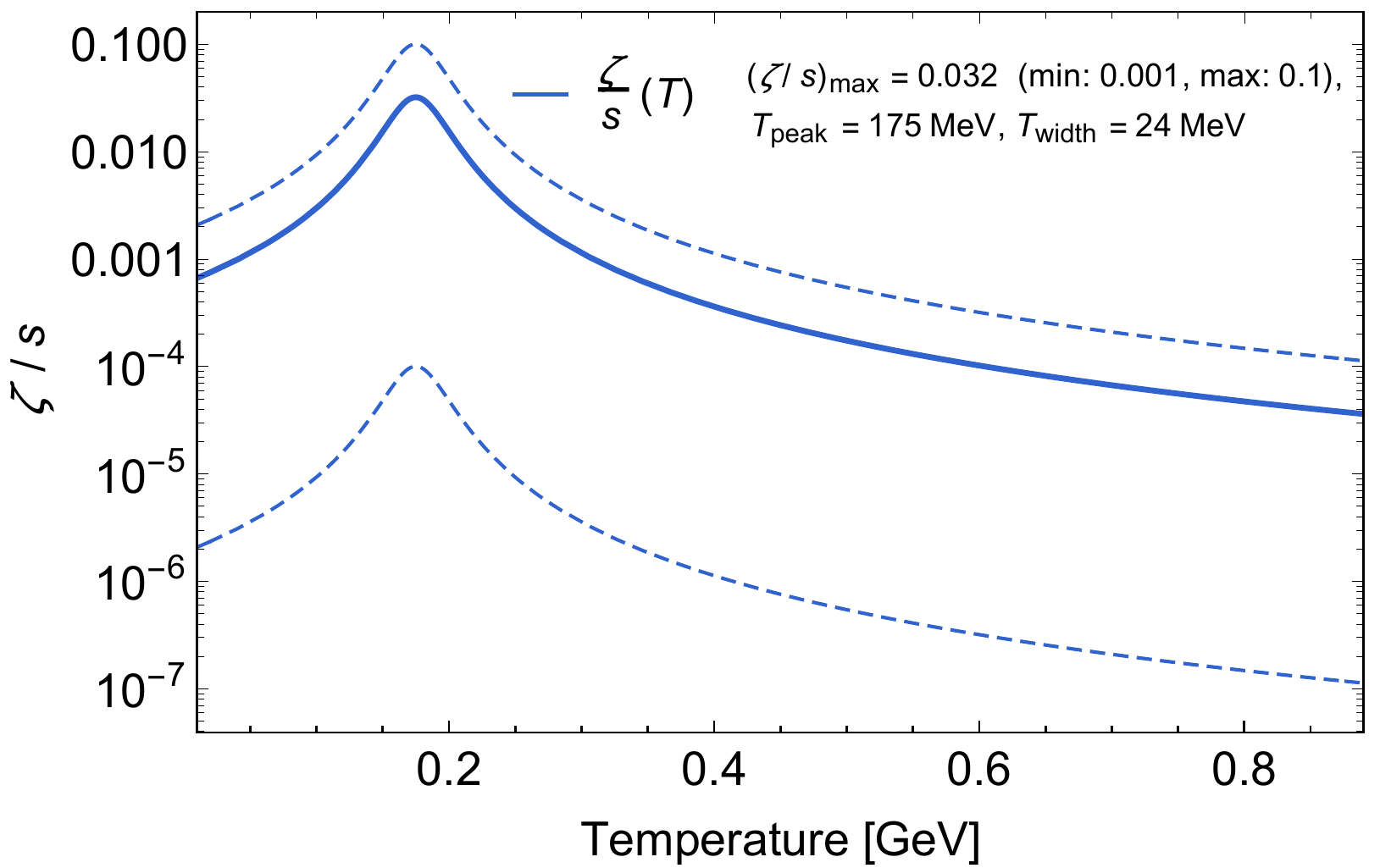}
    \caption{Temperature dependence of the shear (left) and bulk (right) viscosity to entropy ratio as defined in Eq.~\ref{eq:etasQCD}, and~\ref{eq:bulk}, respectively. Both the Yang--Mills and QCD $\eta/s$ distributions are shown, as well as the Kovtun-Son-Starinet (KSS) bound from AdS/CFT calculations, $\eta/s = 1/4\pi$. The dashed blue lines indicate the minimum and maximum values used in the Bayesian analysis for $(\eta/s)_{\rm scale}$ and $\left(\zeta/ s\right)_{\rm max}$.}
    \label{fig:visc_over_entr}
\end{figure*}

We introduce a novel aspect here with respect to Ref.~\cite{Devetak:2019lsk}, by assuming the shear viscosity to entropy ratio $\eta/s$ to be temperature dependent. 
In particular, we exploit the calculation in Yang--Mills theory of Ref.~\cite{Christiansen:2014ypa} with recently updated parameters~\cite{Pawlowski_private}. 
The calculation provides an analytic fit formula describing the temperature dependence of $\eta/s$ as a direct sum of a glueball resonance gas contribution with an algebraic decay at small temperatures and a high-temperature contribution consistent with HTL-resummed perturbation theory. 
The fit function in $\rm SU(3)$ Landau gauge Yang--Mills theory is
\begin{equation}
    \frac{\eta}{s}(T)_{\rm YM} = a\left(\frac{T}{T_{\rm c}} - d\right)^2 + \frac{b}{(T/T_{\rm c})^\delta}.
    \label{eq:etas}
\end{equation}
The first term has been changed with respect to Ref.~\cite{Christiansen:2014ypa} for simplicity since the differences do not play a role for hydrodynamic applications~\cite{Pawlowski_private}. 
The best fit to the full Yang--Mills results is given by the parameters $a=0.0613$, $b=0.00588$, $d=-0.709$, and $\delta=40.3$. 
In the low-temperature regime, the pure glueball resonance gas is not replaced by a hadron resonance gas; the $b$ and $\delta$ parameters are adjusted to $0.02$ and $6.0$ to capture this regime better. 
Finally, an overall correction factor of $4/3$ is used to take the differences in scales and the running couplings in Yang--Mills and QCD into account~\cite{Christiansen:2014ypa}.
On top of this, we add a global scaling $(\eta/s)_{\rm scale}$ as parameter to be estimated with the Bayesian analysis
\begin{equation}
    \frac{\eta}{s}(T)_{\rm QCD} = (\eta/s)_{\rm scale} \cdot \frac{4}{3} \cdot \left[ a\left(\frac{T}{T_{\rm c}} - d\right)^2 + \frac{0.02}{(T/T_{\rm c})^{6}} \right].
    \label{eq:etasQCD}
\end{equation}
The Yang--Mills and QCD $\eta/s$ distributions are presented in the left panel of Fig.~\ref{fig:visc_over_entr}. 
We would like to emphasise that our $\eta/s$ parametrisation is consistently applied throughout both the partonic and hadronic phases.
In contrast, other Bayesian analyses~\cite{Moreland:2018gsh, Bernhard:2019bmu, JETSCAPE:2020shq, JETSCAPE:2020mzn, Nijs:2020ors, Nijs:2020roc, Nijs:2021clz, Nijs:2023yab, Heffernan:2023utr, Liyanage:2023nds, Parkkila:2021tqq, Parkkila:2021yha} commonly adopt a different approach, switching to the UrQMD~\cite{Demir:2008tr} or SMASH~\cite{Rose:2017bjz} transport models at temperature $T_{\rm switch}$ (typically around 145--155 MeV), where $(\eta/s)(T_{\rm switch}) \approx 1$ is employed.

The bulk viscosity to entropy ratio $\zeta / s$ is also considered temperature dependent. 
We assume it to be of the Lorentzian form (shown in the right panel of Fig.~\ref{fig:visc_over_entr})
\begin{equation}
    \frac{\zeta}{s}(T)= \frac{\left(\zeta/s \right)_{\rm max}}{1+\left(\frac{T-T_{\rm peak}}{T_{\rm width}}\right)^2}.
    \label{eq:bulk}
\end{equation}
For now, the peak temperature $T_{\rm peak}$ and width $T_{\rm width}$ are fixed to 175~MeV and 24~MeV, respectively, based on Ref.~\cite{Moreland:2018gsh}. 
The maximum temperature $\left(\zeta/ s\right)_{\rm max}$ is a free parameter for the Bayesian analysis.
We would like to remark that we are aware of the calculations of Refs.~\cite{Karsch:2007jc, Noronha-Hostler:2008kkf} (and its parametrisation in Ref.~\cite{Denicol:2009am}), where the bulk viscosity coefficient shows a maximum near the QCD phase transition temperature $T_{\rm c}$ and starts to decrease almost exponentially while approaching the hadron gas model, as in our assumption.
On the other hand, the temperature dependence of the bulk viscosity in $\rm SU(3)$-gluodynamics is also studied within lattice QCD simulations by Ref.~\cite{Astrakhantsev:2018oue}, which shows that below $T_{\rm c}$, $\zeta/s$ continues to rise.
The same behaviour is observed for a pion gas in Ref.~\cite{Lu:2011df} where the bulk viscosity to entropy ratio rapidly increases at low temperatures.
This rise might be assigned to the rapid decrease of the entropy density below the transition in $\rm SU(3)$-gluodynamics, where the (de)confinement phase transition is of the first order, while in QCD it is a crossover.
A bulk viscosity that keeps increasing as the temperature drops will lead to a large bulk viscous pressure on the freeze-out surface and might bring to some non-physical results during particlisation via the Cooper--Frye procedure.
Since a consensus on the shape of the $\zeta/s$ as a function of the $T$ is not yet available in the literature, we choose to keep using the bulk parametrisation as in Eq.~\ref{eq:bulk} already used in our previous work~\cite{Devetak:2019lsk}\footnote{It is at present also unclear how bulk viscous effects interfere with the chemical freeze-out and a phase of partial chemical equilibrium between chemical and kinetic freeze-out. We plan to revisit this topic in future work.}.
On the freeze-out surface we take the particle distribution function to be given by the equilibrium Bose--Einstein or Fermi--Dirac distribution (depending on the species), modified by additional corrections due to bulk and shear viscous dissipation, 
\begin{equation}
  f = f_\text{eq} + \delta f^\text{bulk} + \delta f^\text{shear}. \label{eq:initialf}
\end{equation}
For the viscous corrections we use the commonly employed parametrisations~\cite{Teaney:2003kp,Paquet:2015lta,Devetak:2019lsk} 
\begin{align}
  &\delta f^\text{bulk} = f_\text{eq}(1\pm f_\text{eq})\left[\frac{\bar E_p}{T}\left(\tfrac{1}{3}-c_s^2\right)-\frac{m^2}{3 T\bar E_p} \right]\frac{\pi_\text{bulk}}{\zeta/\tau_\text{bulk}},\label{eq:bulkini}\\
  &\delta f^\text{shear} = f_\text{eq}(1\pm f_\text{eq})\frac{\pi_{\rho\nu}p^\rho p^\nu}{2(\epsilon+p)T^2}.\label{eq:shearini}
\end{align}
Here $m$ is the mass of the primary resonance.

The bulk and shear relaxation times are taken as derived in Ref.~\cite{Denicol:2014vaa}
\begin{align}
    \frac{\tau_{\rm bulk}}{\zeta/(\epsilon+p)} &= \frac{1}{15\left( \frac{1}{3}- c_s^2 \right)^2 }+\frac{a_{\rm offset}}{\zeta/(\epsilon+p)}, \\
    \frac{\tau_{\rm shear}}{\eta/(\epsilon+p)} &= 
    \begin{cases}
        5 & \text{if } T \geq T_{\rm chem}\\
        5 + (T - T_{\rm chem}) \cdot a_{\rm slope} & \text{if } T < T_{\rm chem}
    \end{cases},
    \label{eq:Taupi}
\end{align}
where $\epsilon$ is the energy density, $p$ is the pressure, $c_s$ is the (temperature dependent) velocity of sound, and $a_{\rm offset}=0.1$~fm$/c$ is a small offset such that a causal evolution of the radial expansion is ensured~\cite{Floerchinger:2017cii}. 
For the same reason, we adjusted the relation between $\eta$ and $\tau_{\rm shear}$ for temperatures $T < T_{\rm chem}$, incorporating an additional $(T - T_{\rm chem}) \cdot a_{\rm slope}$ term with respect to Ref.~\cite{Denicol:2014vaa}. 
This was necessary to ensure that the relaxation time is much larger than the characteristic scale of the hadron resonance gas, for which we expect the scatterings to become more sparse. 
A value of $a_{\rm slope} = 3$~MeV$^{-1}$ was considered for the central analysis after stability checks on the high-\pt region ($\pt>2$~\gevc) of the transverse momentum spectra.

\subsection{Freeze-out and resonance decays: \fastreso}

As the system cools down and dilutes, it changes from a QGP to a hadronic gas. 
Because particle scatterings are no longer efficient in maintaining equilibrium, the fluid dynamic description breaks down, necessitating the conversion of fluid fields to the distribution of hadronic degrees of freedom. 
The Cooper--Frye procedure is used to convert fluid fields to the spectrum of hadron species on a freeze-out surface, which in our work is assumed to be a surface of constant temperature~\cite{Cooper:1974mv}.

In our previous work~\cite{Devetak:2019lsk}, a single freeze-out was considered, where both the chemical and kinetic distributions freeze out simultaneously at temperature $T_{\rm freeze}$.
Here, we improve on this description by introducing a partial chemical equilibrium (PCE) after the chemical freeze-out and before the kinetic freeze-out, i.e.~$T_{\rm kin} < T < T_{\rm chem}$. 
During the PCE, the mean free time for elastic collisions is still smaller than the characteristic expansion time of the expanding fireball, thereby keeping the gas in a state of local kinetic equilibrium. 
The chemical equilibrium is not maintained since the mean free path of the inelastic collisions exceeds this threshold. 

Our description follows the pioneering work of Refs.~\cite{Bebie:1991ij, Huovinen:2007xh}, in which the different particle species in a hadronic gas are treated as being in chemical equilibrium with each other, while the overall gas is not. 
The effective number of long living hadrons, taken in our case as the ones with a characteristic lifetime longer than 10~fm$/c$ (the approximate lifetime of the system~\cite{ALICE:2011dyt}), are fixed at the values they had at the chemical freeze-out. 
The term ``effective'' includes the hadrons which would be produced when all unstable resonances decay, i.e.~$\bar{N}_i = N_i + \sum_j b_j^{i} N_j$. 
Here, $N_i$ is the number of hadrons $i$, $N_j$ the number of resonances $j$, and $b_j^i$ the number of hadrons $i$ formed in the decay of resonance $j$ (including the branching ratio). 
The corresponding effective chemical potential is given in a similar way. 
We then assume the isentropic evolution of ideal hydrodynamics (i.e.~entropy is conserved), meaning that the ratio of effective particle number density ($\bar{n}_i = \bar{N}_i/V$) and entropy density is constant until the kinetic freeze-out. 
From the relation
\begin{equation}
    \frac{\bar{n}_i(T,\mu)}{s(T,\mu)} = \frac{\bar{n}_i(T_{\rm chem}, 0)}{s(T_{\rm chem}, 0)},
    \label{eq:pce}
\end{equation}
one can then obtain $\mu_i(T)$.

On the freeze-out surface we take the particle distribution function to be given by the equilibrium Bose--Einstein or Fermi--Dirac distribution (depending on the species), modified by additional corrections due to bulk and shear viscous dissipation and decays of unstable resonances, as explained in detail in our previous work~\cite{Devetak:2019lsk}. 
For the viscous corrections, we use commonly employed parametrisations~\cite{Teaney:2003kp,Paquet:2015lta}, while the resonance decays are efficiently calculated with the \fastreso package~\cite{Mazeliauskas:2018irt}.
We use a list of approximately 700 resonances from Refs.~\cite{Alba:2017mqu, Alba:2017hhe, Parotto_private}.

\section{Procedure of the Bayesian analysis}
\label{sec:analysis}

As outlined in Sec.~\ref{sec:setup}, our central framework revolves around certain free parameters: the overall normalisation constant ${\rm Norm}$, $(\eta/s)_{\rm min}$\footnote{For convenience, the $(\eta/s)_{\rm scale}$ parameter will be converted from now on to $(\eta/s)_{\rm min}$, representing the minimum of the shear viscosity-to-entropy parametrisation, always located at $T_{\rm min} = 145$~\mev~\cite{Christiansen:2014ypa, Pawlowski_private}.} and $\left(\zeta/ s\right)_{\rm max}$ in the shear and bulk viscosity to entropy ratio parametrisations, the initial fluid time $\tau_0$, and the two freeze-out temperatures $T_{\rm kin}$ and $T_{\rm chem}$.
While ${\rm Norm}$ and $\tau_0$ are considered system-dependent parameters, the others are assumed to converge to the same values for different collision systems and energies.
The primary objective of our Bayesian analysis is to determine these six model parameters simultaneously, allowing them to vary within predefined intervals (see Tab.~\ref{tab:parranges}).
These intervals are based on physical considerations and knowledge from previous studies~\cite{Devetak:2019lsk, Pawlowski_private, Ryu:2017qzn, Andronic:2017pug, ALICE:2013mez}.
It is worth mentioning that we have confirmed \textit{a posteriori} that the optimal values fall within these intervals rather than on their boundaries, and in cases where no clear convergence was obtained, larger intervals were employed.

\begin{table*}[ht!]
  \begin{center}
    \caption{The predefined parameter intervals for the six model parameters across the three collision systems. The ${\rm Norm}$ and $\tau_0$ are considered to be system-dependent parameters. Note that the intervals for the $\tau_0$ parameters for \PbPb \fivenn and \XeXe \fivefourfournn were adjusted \textit{a posteriori}.}
    \begin{tabular}{c|cccccc}
    \hline
    \hline
     & $\left(\zeta/ s\right)_{\rm max}$ & $(\eta/s)_{\rm min}$ & $T_\text{chem}$ (MeV) & $T_\text{kin}$ (MeV) & ${\rm Norm}$ & $\tau_{0}$ (fm/$c$) \\ 
     \hline
     \PbPb ($2.76$~\tev) & & & & & 5--80 & 0.01—3.0 \\
     \PbPb ($5.02$~\tev) & $10^{-4}$--0.3 & 0.08—0.52 & 130--155 & 110--140 & 80--140 & 2.0—7.0 \\
     \XeXe ($5.44$~\tev) & & & & & 70--150 & 2.0—7.0 \\
     \hline
     \hline
    \end{tabular}
  \label{tab:parranges}
  \end{center}
\end{table*}

In contrast to previous Bayesian analyses in the field of heavy-ion physics that employed Gaussian process regression~\cite{Moreland:2018gsh, Bernhard:2019bmu, JETSCAPE:2020shq, JETSCAPE:2020mzn, Nijs:2020ors, Nijs:2020roc, Nijs:2021clz, Nijs:2023yab, Heffernan:2023utr, Liyanage:2023nds}, we introduce a novel approach by utilising neural network emulation.
This paper marks the first application of neural network emulation in this context.
The subsequent two subsections provide a concise introduction to our newly developed framework, which combines neural networks and Markov-Chain Monte Carlo simulations.
For a more comprehensive understanding of both components, we refer the reader to Ref.~\cite{ThesisYannis}, where a detailed overview is presented.

\subsection{Neural Network implementation}
\label{sec:NN}

Although \fluidum is recognised for its very fast execution speeds, the extensive parameter exploration involved in Bayesian analyses necessitates an approach to speed up the simulations.
One common strategy is to train machine learning models to emulate the complete model and utilise these fast emulators within the Markov-Chain Monte Carlo simulations.
In our study, we employ an ensemble of artificial neural networks (ANNs) to emulate the \trento{}\,$+$\,\fluidum{}\,$+$\,\fastreso model.
A neural network is a computational algorithm in the field of supervised learning inspired by the functioning of the brain.
When appropriately constructed and trained, NNs can effectively identify and model complex relationships between inputs and outputs, making them highly promising tools for our research.

In addition to speed, the emulator also needs to exhibit a high level of accuracy.
Achieving this requires careful consideration of the neural network's architecture and the training process.
The training necessitates large datasets to achieve the required accuracy for replacing the simulation outputs.
For each collision system, we use the outputs of ten thousand complete \fluidum{}\,$+$\,\fastreso simulations, with parameters distributed within the ranges presented in Tab.~\ref{tab:parranges}.
The parameter values are generated using Latin hypercube sampling, which ensures a uniform density in an efficient way.
In order to enhance the training performance of the neural networks, we apply a normalisation technique to both the input parameters and the output values, scaling them to the range of $[-1, 1]$.
This procedure leads to a significant increase in the convergence rate and speed of the training process. 
The emulator model incorporates this normalisation, automatically converting the input parameters to the appropriate scale before feeding them through the networks. 
Similarly, the resulting outputs are scaled back to their original ranges.

Simple neural networks provide point predictions without any measure of uncertainty or confidence, which can arise from, e.g., insufficient model complexity or missing information due to unknown data.
Ensemble methods offer an alternative approach where predictions are not solely reliant on a single model; instead, they combine predictions from multiple diverse models within an ensemble.
The predictions, denoted as $f_i$, of the ensemble members $i \in 1, 2, ..., M$ are averaged for the model input $\boldsymbol{x}$,
\begin{equation}
    f_{\rm emu}(\boldsymbol{x}) = \frac{1}{M} \sum^{M}_{i=1} f_i(\boldsymbol{x}),
    \label{eq:EnsembleAvg}
\end{equation}
while the spread among the different ensemble members\footnote{It was tested with a Shapiro--Wilk test that the distribution of the individual neural network predictions can be assumed everywhere to come from a normal distribution~\cite{ThesisYannis}.} and their mean correlation $\rho$ are used to estimate the error on the ensemble prediction via
\begin{equation}
    \sigma_{\rm emu}(\boldsymbol{x}) = \sqrt{\frac{\frac{1}{M} + \frac{M - 1}{M}\rho}{1 - \rho}} \cdot \sqrt{\frac{1}{M} \sum^{M}_{i=1} \left(f_i(\boldsymbol{x}) - f_{\rm emu}(\boldsymbol{x})\right)}
    = c \cdot \hat{\sigma}_{\rm emu}(\boldsymbol{x}).
    \label{eq:EnsembleUnc}
\end{equation}

The correlation among the different neural networks effectively introduces a correction factor, denoted as $c$, to the standard deviation of the neural network predictions $\hat{\sigma}_{\rm emu}(\boldsymbol{x})$.
This correction factor is determined by fitting a $t$-distribution to $(f_{\rm model}(\boldsymbol{x}) - f_{\rm emu}(\boldsymbol{x}))/\hat{\sigma}_{\rm emu}(\boldsymbol{x})$ (here $f_{\rm model}$ represents the original \fluidum\,$+$\,\fastreso output), which should ideally follow a standard normal distribution if the prediction uncertainty accurately captures the prediction error.
We assume that the correlation is independent of the position in parameter space and the considered \pt intervals, allowing the fit to be performed once, considering all configurations $\boldsymbol{x}$ and data points.

For each collision system, we train 100 neural networks using the input data, splitting the data sample into 80\% for training and 20\% for testing and validation purposes.
The PyTorch Python library~\cite{paszke2019pytorch} is employed to construct the neural networks, while the hyperparameters of the neural networks are optimised using a grid search approach facilitated by the Tune Python library~\cite{liaw2018tune}.
The nature of our problem favours shallow neural networks (1--3 hidden layers with $\mathcal{O}(100)$ nodes), a learning rate around 0.01, and the use of a leaky rectified linear unit activation function.

\begin{figure}[tb!]
  \centering
  \includegraphics[width=0.56\linewidth]{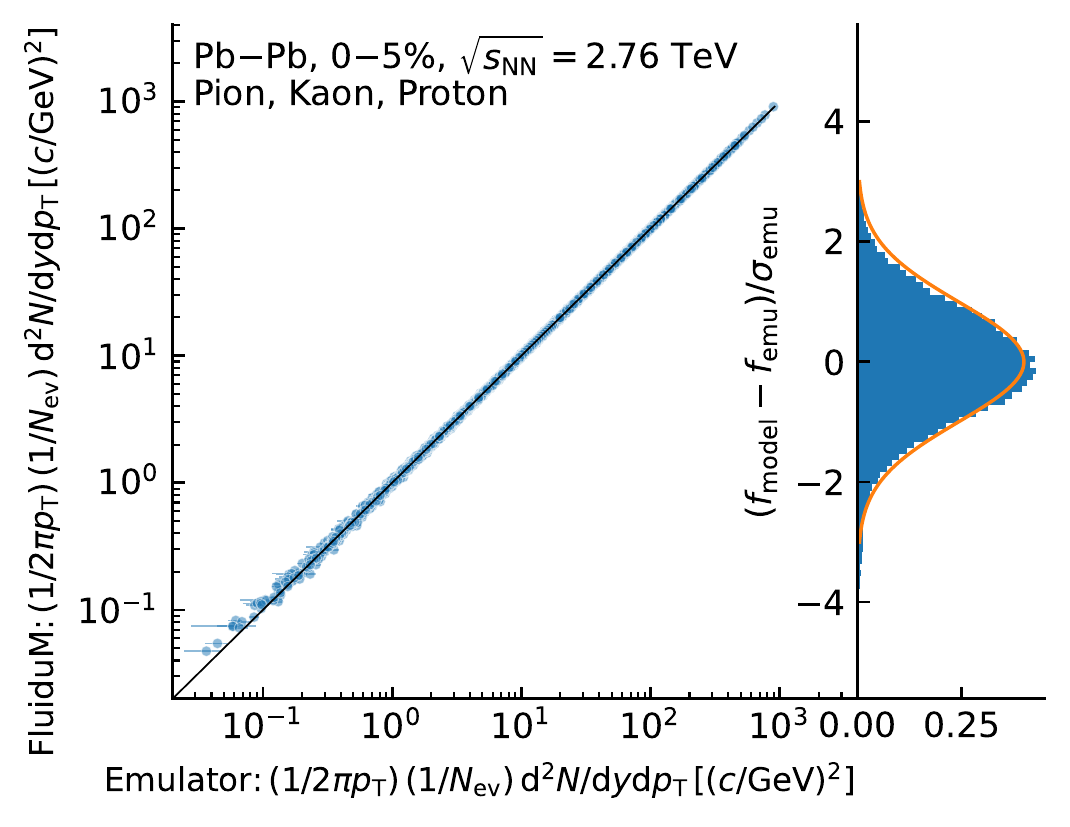}
  \caption{Correlation between the NN ensemble prediction and the original \fluidum{}\,$+$\,\fastreso output for 10 different parameter configurations $\boldsymbol{x}$. Each data point represents the ${\rm d}N^2/{\rm d}y{\rm d}\pt$ value in one experimental \pt interval for either pions, kaons, or protons measured in the 0--5\% most central \PbPb collisions at \twosevensixnn~\cite{ALICE:2013mez}. On the right, the residual distribution is shown in units of the corrected (see text for more details) neural network ensemble uncertainty. A normal distribution with mean zero and standard deviation one is shown in orange to guide the eye.}
  \label{fig:Emulator_uncertainty}
\end{figure}

The neural network ensemble effectively captures the true output of the model simulations, as demonstrated in Fig.~\ref{fig:Emulator_uncertainty}.
The figure presents the correlation between the $(1/2 \pi \pt )(1/N_{\rm ev}) \, {\rm d}^2N/{\rm d}y{\rm d}\pt$ values within experimental \pt intervals for pions, kaons, and protons, as estimated by the emulator model and originally simulated for ten different model parameter configurations $\boldsymbol{x}$.
It is evident that the emulator accurately reproduces the \fluidum{}\,$+$\,\fastreso output, and the emulator uncertainties, which have been appropriately corrected for the leftover correlation among the individual neural networks (as discussed above), effectively capture the residual spread.
Furthermore, our ensemble consisting of 100 NNs exhibits a prediction error, measured in units of the experimental data uncertainty, of $(f_{\rm emu}(\boldsymbol{x}) - f_{\rm model}(\boldsymbol{x}))/\sigma_{\rm exp} = 2.5 \times 10^{-3}$. 
This level of accuracy is considered sufficient for the purposes of this study.
Additionally, our emulator significantly reduces the computation time by a factor $10^4$.

\subsection{Markov-Chain Monte Carlo}
\label{sec:mcmc}

To obtain the posterior probability densities of the $N$ model parameters, we employ Bayesian inference.
The posterior density is inferred from a probabilistic model, and ($N$--1)-dimensional integrals are performed to obtain marginal probability densities for each parameter.
Due to the complexity of our case, an analytic treatment is not feasible. Therefore, we employ the numerical Markov-Chain Monte Carlo (MCMC) method, which is the most efficient approach for exploring the probability space.
In this method, samples are drawn randomly but not independently by constructing a so-called Markov chain.
Each element of the chain is sampled in dependence on its preceding element and only its preceding element.

For our MCMC sampling, we utilise the emcee Python Library~\cite{Foreman-Mackey:2012any}, which implements the affine-invariant ensemble sampler with the so-called ``stretch move''. The logarithmic probability is quantified as the log of the posterior probability
\begin{equation}
    \log{\left(P(\boldsymbol{f}|\boldsymbol{x})\right)} \propto
    \begin{cases}
        -\frac{1}{2} \left[\boldsymbol{f}_{\rm emu}(\boldsymbol{x}) - \boldsymbol{f}_{\rm exp}\right]^{\intercal} \mathbf{\Sigma}^{-1} \left[\boldsymbol{f}_{\rm emu}(\boldsymbol{x}) - \boldsymbol{f}_{\rm exp}\right] & \text{if } x_i^{\rm min} \leq x_i \leq x_i^{\rm max} \, \forall	\, i\\
        -\infty              & \text{otherwise}
    \end{cases}
    ,
    \label{eq:logp}
\end{equation}
Here, $x_i$ represents the $i^{\rm th}$ model input parameter, $x_i^{\rm min}$ and $x_i^{\rm max}$ denote their limits (as defined in Tab.~\ref{tab:parranges}), $\boldsymbol{f}_{\rm exp}$ represents a vector of all the experimental data, $\boldsymbol{f}_{\rm emu}(\boldsymbol{x})$ is the corresponding vector for the emulator model for input $\boldsymbol{x}$, and $\mathbf{\Sigma}$ represents the covariance matrix, which consists of the data and model covariance matrices ($\mathbf{\Sigma} = \mathbf{\Sigma}_{\rm exp} + \mathbf{\Sigma}_{\rm emu}$).
Since no information about the correlations between the experimental uncertainties exist, $\mathbf{\Sigma}_{\rm exp}$ is diagonal with its elements given by the squared sum of the statistical and systematic uncertainties. The model covariance matrix is computed from the ensemble output as
\begin{equation}
    \Sigma_{\rm emu}^{j,k}(\boldsymbol{x}) = c \cdot \frac{1}{M-1} \sum_{i=1}^{M} \left(\left(f_{i}^{j}(\boldsymbol{x}) - f_{\rm emu}^{j}(\boldsymbol{x})\right) \cdot \left(f_{i}^{k}(\boldsymbol{x}) - f_{\rm emu}^{k}(\boldsymbol{x})\right)\right),
    \label{eq:EnsembleCov}
\end{equation}
where the iterators $j$ and $k$ denote specific output values of the vectors.
All entries have to be scaled with the correction factor $c$, as discussed in Sec.~\ref{sec:NN}, to account for the correlation among the different neural networks in the ensemble.
In our MCMC sampling, the prior probability distribution is given by a uniform distribution
\begin{equation}
    p(\boldsymbol{x}) \propto
    \begin{cases}
        1 & \text{if } x_i^{\rm min} \leq x_i \leq x_i^{\rm max} \, \forall	\, i\\
        0 & \text{otherwise}
    \end{cases}
    .
    \label{eq:prior}
\end{equation}

To ensure, that the Markov chain has converged sufficiently, we require the sampling error of the MCMC method to be smaller than 1\%.
The sampling error decreases by $\sqrt{\tau_f / N_{\rm samples}}$, where $\tau_f$ is the integrated autocorrelation time~\cite{ThesisYannis}.
This implies that the product of the number of walkers (chosen to be 300 in our case) and the length of the chains must be greater than $10000\tau_f$.
To prevent too early stopping, we also require that the change of $\tau_f$ (calculated every 100 MCMC steps) is smaller than 1\%.

\begin{figure*}[tb!]
  \centering
  \includegraphics[width=0.9\linewidth]{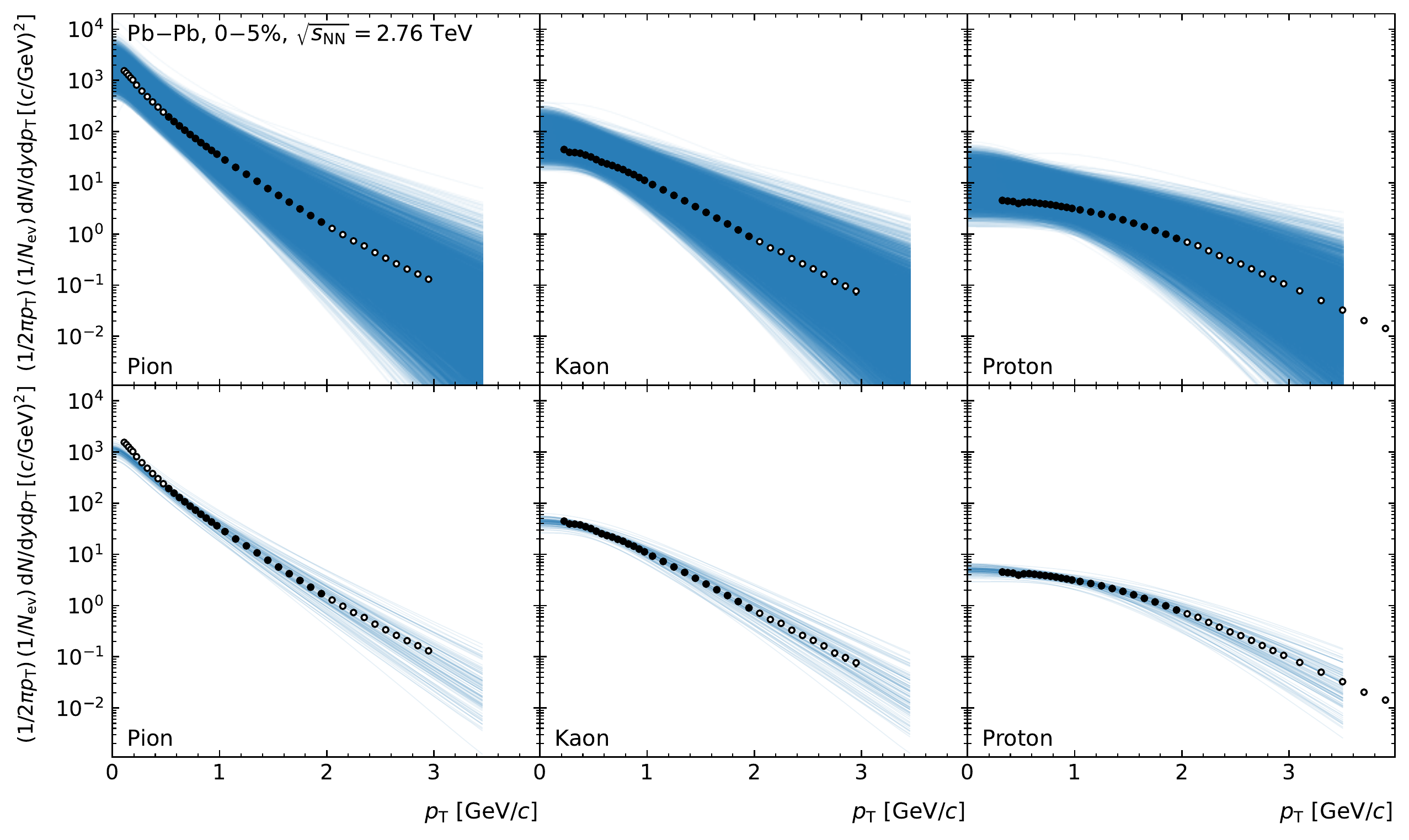}
  \caption{Comparison between simulated (top row; $n=10000$) or emulated (bottom row; $n=100$) \fluidum{}\,$+$\,\fastreso predictions with experimental data for \PbPb collisions at \twosevensixnn in the 0--5\% centrality interval from the ALICE Collaboration~\cite{ALICE:2013mez}. Note that the solid markers show the experimental data used in the MCMC procedure, while the open markers show the full experimental measurement. See text for more details.}
  \label{fig:TrainingPosterior}
\end{figure*}

In order to provide the reader with an initial demonstration of the power of our emulator--MCMC approach, Fig.~\ref{fig:TrainingPosterior} presents a comparison between the simulated and emulated model predictions and the corresponding experimental data for \PbPb collisions at \twosevensixnn~\cite{ALICE:2013mez}.
The top row displays the output of the ten thousand \fluidum{}\,$+$\,\fastreso simulations used for the training of the NNs, and the bottom row presents the predictions generated by the neural-network ensemble emulator using $n=100$ random parameter configurations sampled from the Bayesian posterior distribution obtained through the MCMC chain.
While the training points exhibit a significant spread, which is expected as the parameters are varied over wide ranges (see Tab.~\ref{tab:parranges}), the emulator predictions sampled from the posterior distributions are significantly more constrained and closely follow the experimental data points.

\section{Results}
\label{sec:results}

In this study, we focus on comparing against the experimental measurements of transverse momentum spectra of identified charged hadrons and strange hyperons in the 0--5\% most central \PbPb and \XeXe collisions.
In contrast to traditional Bayesian inference analyses, which often focus on deriving a single `best-fit' scenario, our approach centres on using our framework to assess additional sources of systematic uncertainties, like potential physics that is not included/parametrised in our model, the selection of experimental data, and the treatment of possible unknown correlated uncertainties in the experimental data. 
Through this extensive exploration of the parameter space, we aim to better understand the underlying dynamics of the system.
We emphasise that, in our analysis, all configurations are equally valid, and we do not aim to single out one central result as the definitive outcome. 

The soft particle momentum range $\pt < 2$~\gevc (with variations up to $\pt < 3$~\gevc) always serves as our primary window of investigation, as it is believed to be governed by fluid dynamic approximations of QCD dynamics.
This momentum range is sensitive to important factors such as radial flow, viscous transport coefficients, and the initial conditions of the QGP~\cite{Gale:2013da, Bernhard:2016tnd, Teaney:2009qa, Dubla:2018czx}.
Similar to our previous work~\cite{Devetak:2019lsk}, our default setup excludes pions in the $\pt<0.5$~\gevc range due to the well-known enhancement relative to hydrodynamic simulations~\cite{Gale:2012rq, Nijs:2020roc}.
This low-\pt pion excess is believed to arise from physics features not fully accounted for in hydrodynamic simulations, like: i) Bose--Einstein condensation~\cite{Begun:2015ifa, Begun:2016cva}, ii) increased population of resonances~\cite{Schnedermann:1993ws}, iii) correct treatment of the finite width of $\rho$ meson~\cite{Huovinen:2016xxq}, and iv) effects of critical chiral fluctuations~\cite{Grossi:2021gqi}. 

\subsection{Comparison of different collision system configurations}

\begin{figure*}[tb!]
  \centering
  \includegraphics[width=0.99\linewidth]{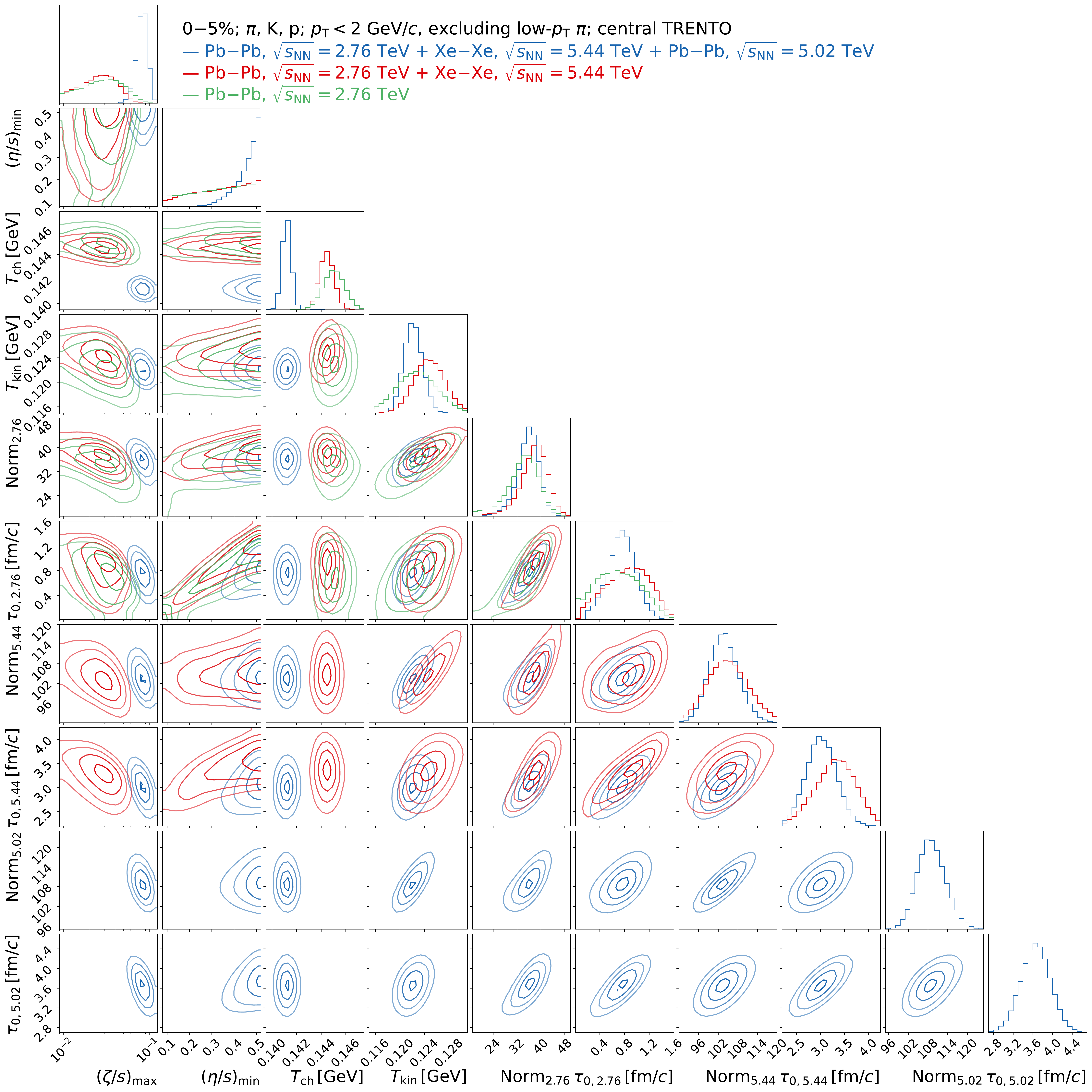}
  \caption{Bayesian posterior distribution of the model input parameters for three different combinations of the collision systems. The diagonal panels show the marginalised distributions of individual model parameters, while the off-diagonal panels present the correlations among pairs of model parameters. See text for more details.}
  \label{fig:MCMCAllSystems}
\end{figure*}

We start by exploring the impact of different collision systems on the constraints of our model parameter.
The Bayesian posterior distributions are presented in Fig.~\ref{fig:MCMCAllSystems}.
The diagonal panels display the marginalised distributions of each individual model parameters, while the off-diagonal panels illustrate the joint distributions for pairs of these parameters marginalised over all others.
The contours represent the (0.5,~1,~1.5,~2)-sigma equivalent regions, encompassing 11.8\%, 39.3\%, 67.5\%, and 86.4\% of the samples.
In addition, the median values and 68\% confidence intervals of the individual model parameters are reported in Tab.~\ref{tab:MedianAllSystems}.
Specifically, we present three configurations: i) the combination of the \PbPb at \twosevensixnn, \PbPb at \fivenn, and \XeXe at \fivefourfournn collision systems in blue; ii) the \PbPb at \twosevensixnn and \XeXe case in red; and iii) only the \twosevensixnn data in green.
The experimental data incorporate the pion ($0.5<\pt<2$~\gevc), kaon ($0.2<\pt<2$~\gevc), and proton ($0.3<\pt<2$~\gevc) $(1/2 \pi \pt )(1/N_{\rm ev}) \, {\rm d}^2N/{\rm d}y{\rm d}\pt$ spectra in the 0--5\% centrality class.
The central \trento configuration is used.

\begin{table}[tb!]
  \begin{center}
    \caption{Posterior parameter estimates corresponding to Fig.~\ref{fig:MCMCAllSystems}. The reported values correspond to the median values and 68\% confidence intervals.}
    \begin{tabular}{l|c|c|c}
    \hline
    \hline
     & 2.76 -- 5.44 -- 5.02~\tev & 2.76 -- 5.44~\tev & 2.76~\tev \\[1pt]
    \hline 
    $\left(\zeta/ s\right)_{\rm max}$ & $0.083^{+0.012}_{-0.012}$ & $0.022^{+0.016}_{-0.014}$ & $0.023^{+0.022}_{-0.016}$ \\[2pt]
    $(\eta/s)_{\rm min}$ & unconstr. & unconstr. & unconstr. \\[2pt]
    $T_\text{chem}$ [MeV] & $141^{+0}_{-0}$ & $144^{+1}_{-1}$ & $145^{+1}_{-1}$ \\[2pt]
    $T_\text{kin}$ [MeV] & $122^{+2}_{-1}$ & $125^{+3}_{-2}$ & $123^{+3}_{-3}$ \\[2pt]
    ${\rm Norm}_{\rm 2.76}$ & $36.0^{+3.1}_{-3.6}$ & $38.2^{+3.8}_{-4.3}$ & $34.6^{+4.5}_{-5.3}$ \\[2pt]
    $\tau_{0,2.76}$ [fm/$c$] & $0.76^{+0.21}_{-0.23}$ & $0.87^{+0.34}_{-0.40}$ & $0.69^{+0.39}_{-0.37}$ \\[2pt]
    ${\rm Norm}_{\rm 5.44}$ & $103.7^{+4.6}_{-4.4}$ & $104.9^{+6.7}_{-6.2}$ & n/a \\[2pt]
    $\tau_{0,5.44}$ [fm/$c$] & $3.01^{+0.31}_{-0.31}$ & $3.32^{+0.40}_{-0.44}$ & n/a \\[2pt]
    ${\rm Norm}_{\rm 5.02}$ & $109.0^{+4.5}_{-4.1}$ & n/a & n/a \\[2pt]
    $\tau_{0,5.02}$ [fm/$c$] & $3.66^{+0.30}_{-0.30}$ & n/a & n/a \\[2pt]
    \hline
    \hline
    \end{tabular}
  \label{tab:MedianAllSystems}
  \end{center}
\end{table}

Remarkably, even with a limited number of \pt spectra observables, we find that most model parameters are well constrained, underscoring the potential of Bayesian analyses in providing valuable insights into the properties of the QGP.
The one parameter that appears to prefer higher values beyond the upper bound of the allowed interval is the minimum value of the shear viscosity to entropy ratio parametrisation. 
We attribute this behaviour to the limited sensitivity of the current observables to the shear viscosity of the system.
Future work incorporating experimental flow data is expected to further address this point.
However, we emphasise that our $\eta/s$ parametrisation is consistently applied throughout both the partonic and hadronic phases, possibly leading to higher minimum values compared to other Bayesian analyses, where an afterburner is used for the hadronic phase~\cite{Moreland:2018gsh, Bernhard:2019bmu, JETSCAPE:2020mzn, Nijs:2020ors, Nijs:2020roc, Nijs:2021clz, Nijs:2023yab, Heffernan:2023utr, Liyanage:2023nds, Parkkila:2021tqq, Parkkila:2021yha}.

When comparing the three scenarios, most parameters are compatible within their uncertainties (which naturally decrease when more experimental data is incorporated).
However, noticeable differences emerge, particularly in the maximum temperature for the bulk viscosity to entropy ratio and the chemical freeze-out temperature, which appear to be influenced by the inclusion of the \PbPb at \fivenn data; the experimental data which we observed to be the most difficult to fit.
Since no immediate physical explanation is available, we recommend further investigation in a future analysis, especially as additional experimental observables are included.
Another notable distinction is observed in the ${\rm Norm}$ and $\tau_0$ parameters for the various collision systems, with a substantial increase between the \PbPb at \twosevensixnn and the other two systems. 

\subsection{Effect of modified initial conditions and inclusion of strange hyperons}

\begin{figure*}[tb!]
  \centering
  \includegraphics[width=.7\linewidth]{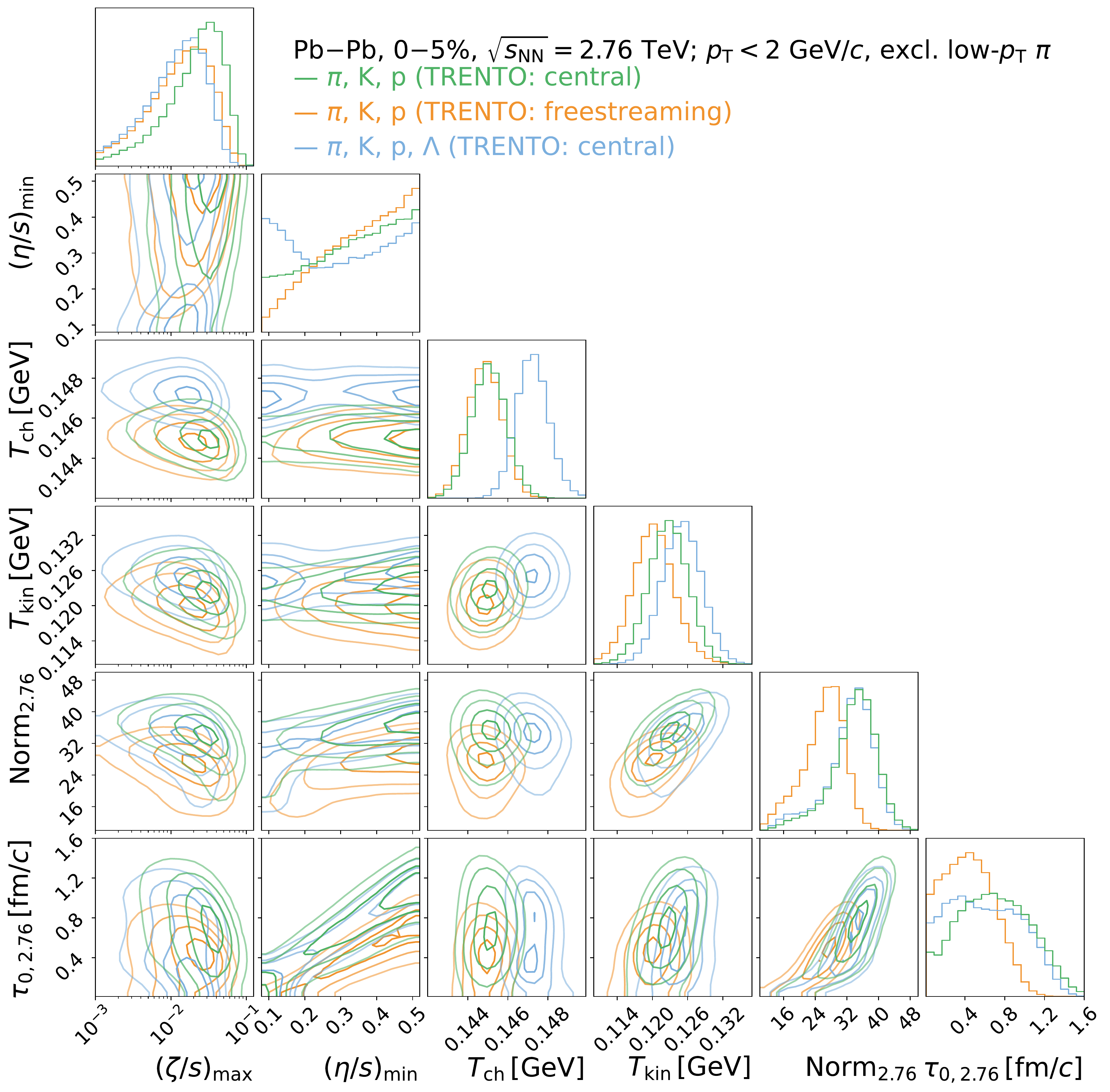}
  \caption{Bayesian posterior distribution of the model input parameters for three different configurations exploiting the \twosevensixnn \PbPb experimental data. The diagonal panels show the marginalised distributions of individual model parameters, while the off-diagonal panels present the correlations among pairs of model parameters. See text for more details.}
  \label{fig:MCMCTRENTOLambda}
\end{figure*}

Shifting the focus to two other crucial aspects of our analysis, we transition to Fig.~\ref{fig:MCMCTRENTOLambda}, where we explore the implications of the following scenarios: i) a modification of the initial \trento configuration to mimic a free-streaming phase, as discussed in Sec.~\ref{sec:initialcond}; and ii) the inclusion of the experimental $(1/2 \pi \pt )(1/N_{\rm ev})\, {\rm d}^2N/{\rm d}y{\rm d}\pt$ data of the strange hyperon $\Lambda$ ($0.6<\pt<2$~\gevc).
We compare the resulting posterior distributions with the single \PbPb \twosevensixnn case discussed earlier. 

When employing the modified \trento settings to mimic the free-streaming phase, we observe a notable decrease in the values of the ${\rm Norm}$ and $\tau_0$ parameters.
This decrease can be attributed to the larger integrated transverse density $\int {\rm d}^2x T_{\rm R}(\vec x)$ in this configuration.
Because of the relationship expressed by Eq.~\ref{eq:sinitial}, there exists a correlation between the magnitudes of the ${\rm Norm}$ and $\tau_0$ parameters.
Consequently, interpreting the decrease of $\tau_0$ in a physical context becomes challenging, and the smaller value indicates only partly the (artificial) smoothening out of the initial entropy density profile by reducing its ``spikiness''. 

Furthermore, the inclusion of the $\Lambda$ hyperon data leads to an increase in the chemical freeze-out temperature.
This finding is consistent with previous observations using hydrodynamic simulations~\cite{Devetak:2019lsk, Ryu:2017qzn}, suggesting that strange and multi-strange baryons are more sensitive to changes in the transition temperature between the fluid evolution and the hadronic transport phases.
This aligns with proposals in the literature indicating that strange hadrons may undergo chemical freeze-out earlier than non-strange particles~\cite{vanHecke:1999jh, Arbex:2001vx, Chatterjee:2013yga, Bellwied:2013cta}. 

\subsection{Exploring variations in the analysis setup}

\begin{figure*}[tb!]
  \centering
  \includegraphics[width=0.99\linewidth]{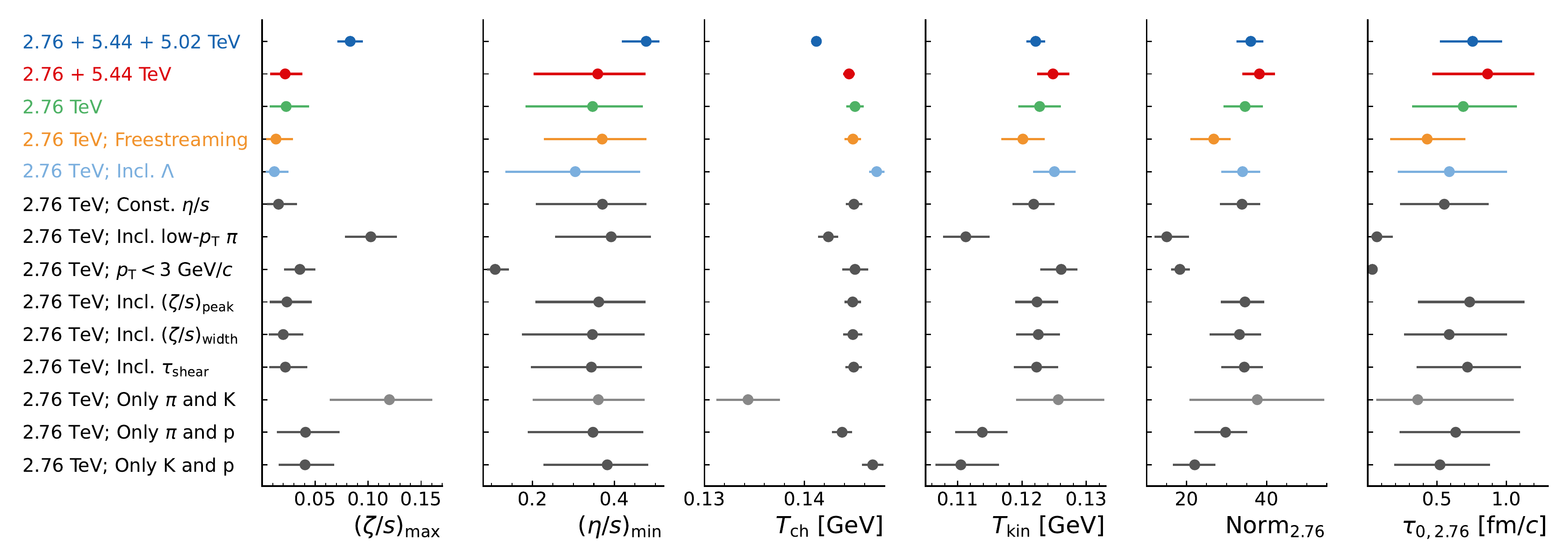}
  \caption{The median values and 68\% confidence intervals of the marginalised Bayesian posterior distributions for each model parameter across all analysed configurations. Note that due to the collision system dependency, the ${\rm Norm}$ and $\tau_0$ parameters for \XeXe and \PbPb at \fivenn are reported only in Tab.~\ref{tab:MedianAllSystems}. The configuration involving only pions and kaons is depicted in a lighter colour, reflecting the lack of convergence to a single minimum in the MCMC procedure for this specific setup.}
  \label{fig:MCMCSummary}
\end{figure*}

Figure~\ref{fig:MCMCSummary} summarises the previously discussed variations by reporting the median values and 68\% confidence intervals of the marginalised Bayesian posterior distributions for each model parameter.
Several smaller variations, exclusively applied to the \PbPb collision system at \twosevensixnn, are included as well.
We assessed the effect of: i) using a constant shear viscosity to entropy ratio instead of the temperature-dependent version; ii) including pions with transverse momentum below 500~\mevc; iii) increasing the transverse momentum limit to 3~\gevc for all particle species; iv--vi) including $(\zeta/s)_{\rm peak}$ (150--200~\mev), $(\zeta/s)_{\rm width}$ (10--100~\mev), or $\tau_{\rm shear}$ ($a_{\rm slope} = 0$--$10$~MeV$^{-1}$) as seventh model variable in the MCMC procedure; and vii--ix) considering only pions and kaons, pions and protons, and kaons and protons.

Overall, the values of the extracted model parameters demonstrate a reasonable stability across all configurations.
When considering a constant shear viscosity to entropy ratio, the results remain consistent with those obtained from the temperature-dependent formulation (Eq.~\ref{eq:etas}).
The observation that the constant value for $\eta/s$ closely resembles $(\eta/s)_{\rm min}$ in the temperature-dependent parametrisation suggests that, in the latter case, we are primarily sensitive to the region close to the phase transition.
The inclusion of low transverse momentum pions impacts several model variables.
Specifically, it leads to a substantially smaller $\tau_0$, marginally lower freeze-out temperatures, and a slightly larger $(\zeta/s)_{\rm max}$.
Given the expected non-hydrodynamic origin of the enhancement in the low-\pt pion spectra~\cite{Begun:2015ifa, Begun:2016cva, Schnedermann:1993ws, Huovinen:2016xxq, Grossi:2021gqi}, this underscores the importance of restricting the analysis within the range where hydrodynamics is expected to be applicable.
Increasing the upper \pt limit, instead, yields a notable impact on the $(\eta/s)_{\rm min}$ value.
This parameter now tends to favour lower values, nearing the theoretical lower bound of $1/4\pi$ derived from AdS/CFT.
This behaviour likely arises from the shear correction in the computation of the final particle spectrum that scales with $\pt^2$~\cite{Mazeliauskas:2018irt, Kirchner:2023fsj}, i.e.~expanding the analysed \pt interval includes points that are much more sensitive to the shear corrections and we would thus expect a change in the $(\eta/s)_{\rm min}$ parameter.

The introduction of $(\zeta/s)_{\rm peak}$, $(\zeta/s)_{\rm width}$, or $a_{\rm slope}$ as the seventh model variable in the MCMC procedure has a negligible impact on the values of the original six parameters.
These parameters remain consistent with those obtained from the central \PbPb \twosevensixnn configuration.
However, it is important to note that none of the posterior distributions for these seventh model parameters converge; all three distributions touch the upper limit of their respective intervals.
This emphasises the need for first-principle calculations of the bulk viscosity which could replace the currently assumed Lorentzian form.
Finally, excluding one of the three particle species (pions, kaons, or protons) results in small changes in the ${\rm Norm}$ and freeze-out temperatures, but still compatible within 1--2$\sigma$.
When considering only pions and kaons, the posterior distributions exhibit a dual structure, where one component aligns well with the central \PbPb \twosevensixnn configuration, while the other shows significantly different values.
Due to this configuration's lack of convergence, we choose to exclude it from further analysis. 

\subsection{Discussion and comparison to experimental data}

We start the discussion by comparing our extracted model parameters in Fig.~\ref{fig:MCMCSummary} with those obtained from analogous Bayesian inference analyses.
The values for $(\zeta/s)_{\rm max}$ are in agreement with analyses employing the same functional form~\cite{Moreland:2018gsh, Bernhard:2019bmu, Nijs:2021clz, Nijs:2023yab, Parkkila:2021tqq, Parkkila:2021yha}, and compatible with~\cite{Liyanage:2023nds} or lower by 2--3$\sigma$~\cite{JETSCAPE:2020shq, JETSCAPE:2020mzn, Heffernan:2023utr} when compared to Bayesian analyses that introduce a slight adjustment to allow for an asymmetry in temperature around the peak.
Regarding the $(\eta/s)_{\rm min}$ parameter, all other Bayesian analyses typically find values around 0.10~\cite{Moreland:2018gsh, Bernhard:2019bmu, JETSCAPE:2020shq, JETSCAPE:2020mzn, Nijs:2020ors, Nijs:2020roc, Nijs:2021clz, Nijs:2023yab, Heffernan:2023utr, Liyanage:2023nds, Parkkila:2021tqq, Parkkila:2021yha}, while our posterior distributions hint at values beyond the upper bound of 0.52.
We remind the reader that we attribute this behaviour to two factors: i) a limited sensitivity of the current observables to the shear viscosity of the system, and ii) a different strategy regarding the hadronic phase of the system.
On account of reason ii), a direct comparison of our extracted freeze-out temperatures with those from other Bayesian analyses seems unfeasible.
These analyses typically employ a single freeze-out temperature $T_{\rm switch}$ (converging to values between 130 and 160~\mev), while their hadronic afterburner continues to evolve until yields and momentum distributions cease changing.
However, since most of the hadronic yields vary by less than 20\% as a consequence of inelastic collisions in the afterburner phase, the switching and chemical freeze-out temperature are typically associated with each other~\cite{Song:2010aq}.
In this context, it is noteworthy that we obtain values that are compatible with the switching temperatures found in Refs.~\cite{Heffernan:2023utr, Liyanage:2023nds}, 5--15~\mev lower with respect to Refs.~\cite{Nijs:2020ors, Nijs:2020roc, Nijs:2021clz, Nijs:2023yab, Parkkila:2021tqq, Parkkila:2021yha}, and slightly higher values than Refs.~\cite{JETSCAPE:2020shq, JETSCAPE:2020mzn}.
Instead, the statistical hadronisation model of Ref.~\cite{Andronic:2017pug} suggests chemical freeze-out temperatures of approximately 10~\mev higher, while blast-wave fits to the \pt distributions of identified hadrons in central collisions estimate kinematic freeze-out temperatures of about 20~\mev lower than ours~\cite{ALICE:2013mez}.
The extracted $\tau_0$ values for the \PbPb \twosevensixnn system align roughly with the time of the prehydrodynamic phase in the other Bayesian analyses, typically spanning 0.3--1.0~fm$/c$~\cite{Moreland:2018gsh, Bernhard:2019bmu, JETSCAPE:2020shq, JETSCAPE:2020mzn, Nijs:2020ors, Nijs:2020roc, Nijs:2021clz, Nijs:2023yab, Heffernan:2023utr, Parkkila:2021tqq, Parkkila:2021yha}. 
However, due to the correlation with the ${\rm Norm}$ parameter (as depicted in Eq.~\ref{eq:sinitial}), our consideration of this parameter as system dependent (resulting in significantly larger values for the \XeXe and \PbPb \fivenn collision systems), and the absence of a free-streaming phase in our setup, we would refrain from interpreting them as being the same parameter in a physical sense. 

\begin{figure*}[tb!]
  \centering
  \includegraphics[width=0.99\linewidth]{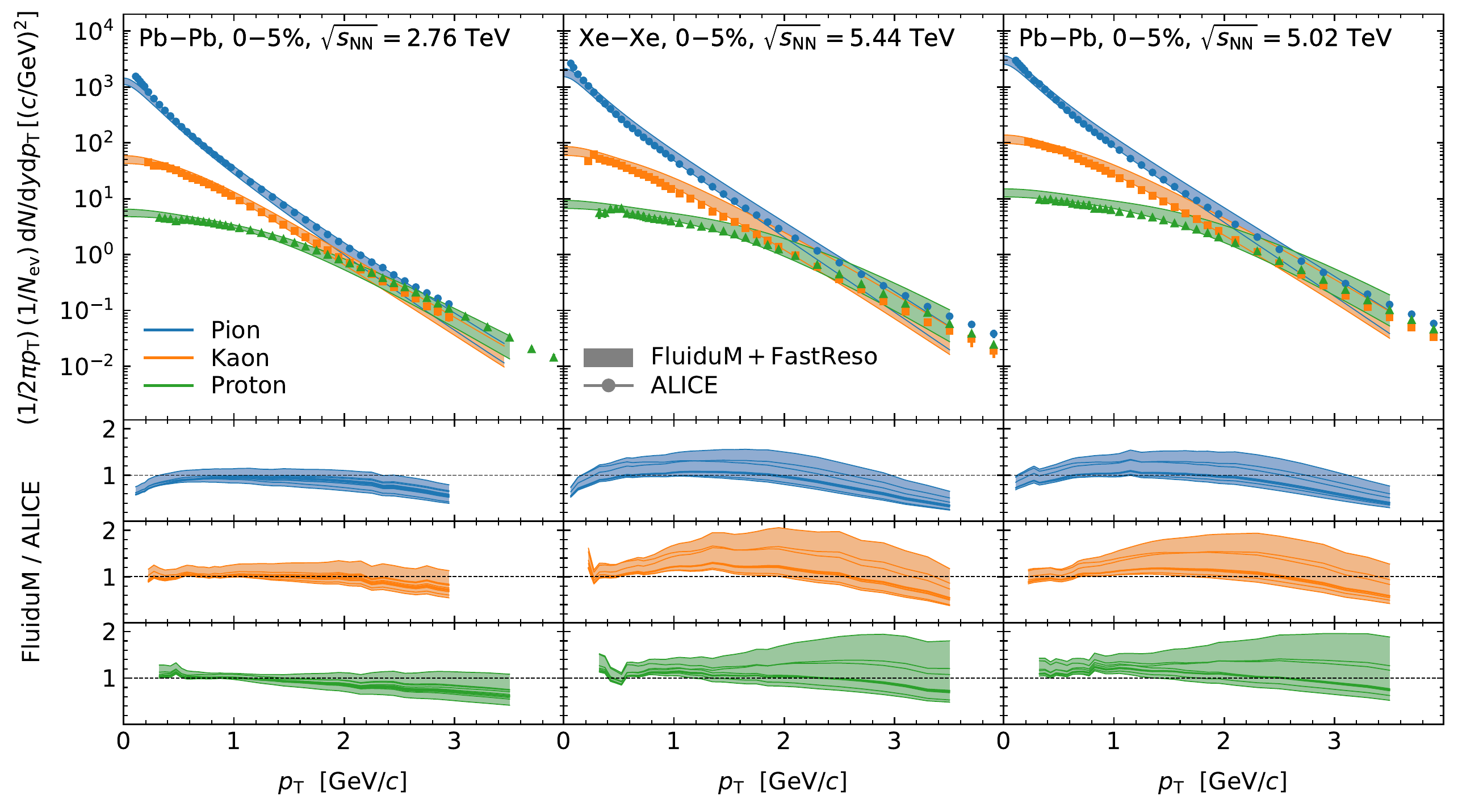}
  \caption{The top panels display the transverse momentum spectra for pions, kaons, and protons in the 0--5\% centrality class for the three collision systems. The spectra as simulated with the \fluidum{}\,$+$\,\fastreso framework, using the extracted model parameters from our Bayesian analysis (see Fig.~\ref{fig:MCMCSummary}), are compared with experimental data from the ALICE Collaboration~\cite{ALICE:2013mez,ALICE:2019hno,ALICE:2021lsv}. The bottom panels present the respective ratios between theory and experimental data for each hadron species. Note the experimental uncertainties are not taken into account in the ratio.}
  \label{fig:FluiduMvsALICE}
\end{figure*}

In Fig.~\ref{fig:FluiduMvsALICE}, we translate the values of our model parameters from Fig.~\ref{fig:MCMCSummary} into a final \fluidum{}\,$+$\,\fastreso prediction for the pion, kaon, and proton $(1/2 \pi \pt )(1/N_{\rm ev}) \, {\rm d}^2N/{\rm d}y{\rm d}\pt$ spectra in the 0--5\% centrality class for the three collision systems.
For the analysis configurations for which the MCMC procedure was only run for the \PbPb \twosevensixnn system, we use the ${\rm Norm}$ and $\tau_0$ from Tab.~\ref{tab:MedianAllSystems} for the \PbPb at \fivenn and \XeXe at \fivefourfournn calculations.
The theoretical uncertainty is estimated as the full envelope of the various trials\footnote{Excluding the ``$\pt < 3$~\gevc'' and ``Only $\pi$ and K'' cases as previously discussed.}.
The simulations exhibit good quantitative agreement with experimental measurements.
The theoretical uncertainties are approximately 20\% for the \PbPb \twosevensixnn system and about 40\% for the \XeXe and \fivenn \PbPb systems.
This difference can be attributed to the fact that we primarily explored variations of the analysis configuration using the \PbPb \twosevensixnn system. 
In future work, where we will include anisotropic flow observables and explore additional centrality classes, we intend to conduct systematic variations for all three systems simultaneously.
We expect that the resulting theoretical uncertainties will be of comparable magnitude for each system.

\section{Summary}
\label{sec:summary}

In summary, we presented a Bayesian analysis employing our \trento{}\,$+$\,\fluidum{}\,$+$\,\fastreso framework to determine key parameters of the QGP. These parameters included the shear and bulk viscosity to entropy ratios, the initialisation time, the initial entropy density, and the freeze-out temperatures. 

Modern machine learning tools were utilised to perform a global search in multidimensional space to extract the posterior distributions of the model parameters. In contrast to previous Bayesian analyses employing Gaussian process regression, this study pioneered the use of neural network ensemble emulation. This innovation offered several computational advantages, including significantly reduced training time, lower memory usage, the ability to handle any number of inputs and outputs, and a rigorous uncertainty determination. 

Another notable improvement in this work concerns our theoretical framework. Instead of a single freeze-out, we now separate the chemical and kinematic freeze-outs, incorporating the partial chemical equilibrium to describe the later stages of the evolution. In addition, we replaced the use of a constant value for the shear viscosity to entropy ratio with a Yang--Mills theory-based parametrisation.

Our theoretical model is compared against experimental measurements of transverse momentum spectra for identified charged hadrons ($\pi$, $\rm K$, $\rm p$) and strange hyperons ($\Lambda$) in \PbPb collisions at \twosevensixnn and \fivenn, as well as \XeXe collisions at \fivefourfournn, from the ALICE Collaboration. Our focus lays on the 0--5\% centrality class, where the \fluidum background--fluctuation splitting ansatz works best. Furthermore, the restriction to only one centrality class was imposed such that we are able to keep the initial-state parameters, which play an important role in the centrality dependence of the observables, fixed. In future work, we will include anisotropic flow observables and explore additional centrality classes.

Thanks to the computational efficiency of our framework, we could employ it to assess systematic uncertainties by varying key components of the analysis. In other words, we did not focus on a single `best-fit' scenario but conducted an extensive exploration of the parameter space to better understand the underlying dynamics of the system. Multiple variations, including data from various collision systems, modified initial conditions, inclusion and exclusion of hadron species, variations in the shear and bulk viscosity to entropy ratios, and the use of different \pt ranges, were performed. Overall, the extracted model parameters exhibit a reasonable stability across the configurations. Furthermore, our results are in agreement with previous Bayesian analyses, except for the $(\eta/s)_{\rm min}$, which we attributed to limited sensitivity of the observables and a different theoretical strategy concerning the hadronic phase.

Finally, we translated our extracted parameters into \fluidum+\fastreso predictions for the pion, kaon, and proton $(1/2 \pi \pt )(1/N_{\rm ev}) \, {\rm d}^2N/{\rm d}y{\rm d}\pt$ spectra in the 0--5\% centrality class for the three collision systems. They demonstrate strong quantitative agreement with experimental measurements from the ALICE Collaboration.

\section*{Acknowledgements}

The authors wish to thank Christian Sonnabend and Lukas Kreis for their contribution in the early phases of this work.
This work is part of and supported by the DFG Collaborative Research Centre ``SFB 1225 (ISOQUANT)''.
A.D. is partially supported by the Netherlands Organisation for Scientific Research (NWO) under the grant 19DRDN011, VI.Veni.192.039.
Computational resources have been provided by the GSI Helmholtzzentrum f{\"u}r  Schwerionenforschung.

\bibliography{apssamp}

\begin{thebibliography}{82}%
\makeatletter
\providecommand \@ifxundefined [1]{%
 \@ifx{#1\undefined}
}%
\providecommand \@ifnum [1]{%
 \ifnum #1\expandafter \@firstoftwo
 \else \expandafter \@secondoftwo
 \fi
}%
\providecommand \@ifx [1]{%
 \ifx #1\expandafter \@firstoftwo
 \else \expandafter \@secondoftwo
 \fi
}%
\providecommand \natexlab [1]{#1}%
\providecommand \enquote  [1]{``#1''}%
\providecommand \bibnamefont  [1]{#1}%
\providecommand \bibfnamefont [1]{#1}%
\providecommand \citenamefont [1]{#1}%
\providecommand \href@noop [0]{\@secondoftwo}%
\providecommand \href [0]{\begingroup \@sanitize@url \@href}%
\providecommand \@href[1]{\@@startlink{#1}\@@href}%
\providecommand \@@href[1]{\endgroup#1\@@endlink}%
\providecommand \@sanitize@url [0]{\catcode `\\12\catcode `\$12\catcode
  `\&12\catcode `\#12\catcode `\^12\catcode `\_12\catcode `\%12\relax}%
\providecommand \@@startlink[1]{}%
\providecommand \@@endlink[0]{}%
\providecommand \url  [0]{\begingroup\@sanitize@url \@url }%
\providecommand \@url [1]{\endgroup\@href {#1}{\urlprefix }}%
\providecommand \urlprefix  [0]{URL }%
\providecommand \Eprint [0]{\href }%
\providecommand \doibase [0]{https://doi.org/}%
\providecommand \selectlanguage [0]{\@gobble}%
\providecommand \bibinfo  [0]{\@secondoftwo}%
\providecommand \bibfield  [0]{\@secondoftwo}%
\providecommand \translation [1]{[#1]}%
\providecommand \BibitemOpen [0]{}%
\providecommand \bibitemStop [0]{}%
\providecommand \bibitemNoStop [0]{.\EOS\space}%
\providecommand \EOS [0]{\spacefactor3000\relax}%
\providecommand \BibitemShut  [1]{\csname bibitem#1\endcsname}%
\let\auto@bib@innerbib\@empty
\bibitem [{\citenamefont {Busza}\ \emph {et~al.}(2018)\citenamefont {Busza},
  \citenamefont {Rajagopal},\ and\ \citenamefont {van~der
  Schee}}]{Busza:2018rrf}%
  \BibitemOpen
  \bibfield  {author} {\bibinfo {author} {\bibfnamefont {W.}~\bibnamefont
  {Busza}}, \bibinfo {author} {\bibfnamefont {K.}~\bibnamefont {Rajagopal}},\
  and\ \bibinfo {author} {\bibfnamefont {W.}~\bibnamefont {van~der Schee}},\
  }\href {https://doi.org/10.1146/annurev-nucl-101917-020852} {\bibfield
  {journal} {\bibinfo  {journal} {Ann. Rev. Nucl. Part. Sci.}\ }\textbf
  {\bibinfo {volume} {68}},\ \bibinfo {pages} {339} (\bibinfo {year} {2018})},\
  \Eprint {https://arxiv.org/abs/1802.04801} {arXiv:1802.04801 [hep-ph]}
  \BibitemShut {NoStop}%
\bibitem [{\citenamefont {Aamodt}\ \emph {et~al.}(2010)\citenamefont {Aamodt}
  \emph {et~al.}}]{ALICE:2010suc}%
  \BibitemOpen
  \bibfield  {author} {\bibinfo {author} {\bibfnamefont {K.}~\bibnamefont
  {Aamodt}} \emph {et~al.} (\bibinfo {collaboration} {ALICE}),\ }\href
  {https://doi.org/10.1103/PhysRevLett.105.252302} {\bibfield  {journal}
  {\bibinfo  {journal} {Phys. Rev. Lett.}\ }\textbf {\bibinfo {volume} {105}},\
  \bibinfo {pages} {252302} (\bibinfo {year} {2010})},\ \Eprint
  {https://arxiv.org/abs/1011.3914} {arXiv:1011.3914 [nucl-ex]} \BibitemShut
  {NoStop}%
\bibitem [{\citenamefont {Adams}\ \emph {et~al.}(2005)\citenamefont {Adams}
  \emph {et~al.}}]{STAR:2005gfr}%
  \BibitemOpen
  \bibfield  {author} {\bibinfo {author} {\bibfnamefont {J.}~\bibnamefont
  {Adams}} \emph {et~al.} (\bibinfo {collaboration} {STAR}),\ }\href
  {https://doi.org/10.1016/j.nuclphysa.2005.03.085} {\bibfield  {journal}
  {\bibinfo  {journal} {Nucl. Phys. A}\ }\textbf {\bibinfo {volume} {757}},\
  \bibinfo {pages} {102} (\bibinfo {year} {2005})},\ \Eprint
  {https://arxiv.org/abs/nucl-ex/0501009} {arXiv:nucl-ex/0501009} \BibitemShut
  {NoStop}%
\bibitem [{\citenamefont {Adcox}\ \emph {et~al.}(2005)\citenamefont {Adcox}
  \emph {et~al.}}]{PHENIX:2004vcz}%
  \BibitemOpen
  \bibfield  {author} {\bibinfo {author} {\bibfnamefont {K.}~\bibnamefont
  {Adcox}} \emph {et~al.} (\bibinfo {collaboration} {PHENIX}),\ }\href
  {https://doi.org/10.1016/j.nuclphysa.2005.03.086} {\bibfield  {journal}
  {\bibinfo  {journal} {Nucl. Phys. A}\ }\textbf {\bibinfo {volume} {757}},\
  \bibinfo {pages} {184} (\bibinfo {year} {2005})},\ \Eprint
  {https://arxiv.org/abs/nucl-ex/0410003} {arXiv:nucl-ex/0410003} \BibitemShut
  {NoStop}%
\bibitem [{\citenamefont {Acharya}\ \emph {et~al.}(2019)\citenamefont {Acharya}
  \emph {et~al.}}]{ALICE:2019zfl}%
  \BibitemOpen
  \bibfield  {author} {\bibinfo {author} {\bibfnamefont {S.}~\bibnamefont
  {Acharya}} \emph {et~al.} (\bibinfo {collaboration} {ALICE}),\ }\href
  {https://doi.org/10.1103/PhysRevLett.123.142301} {\bibfield  {journal}
  {\bibinfo  {journal} {Phys. Rev. Lett.}\ }\textbf {\bibinfo {volume} {123}},\
  \bibinfo {pages} {142301} (\bibinfo {year} {2019})},\ \Eprint
  {https://arxiv.org/abs/1903.01790} {arXiv:1903.01790 [nucl-ex]} \BibitemShut
  {NoStop}%
\bibitem [{\citenamefont {Aidala}\ \emph {et~al.}(2019)\citenamefont {Aidala}
  \emph {et~al.}}]{PHENIX:2018lia}%
  \BibitemOpen
  \bibfield  {author} {\bibinfo {author} {\bibfnamefont {C.}~\bibnamefont
  {Aidala}} \emph {et~al.} (\bibinfo {collaboration} {PHENIX}),\ }\href
  {https://doi.org/10.1038/s41567-018-0360-0} {\bibfield  {journal} {\bibinfo
  {journal} {Nature Phys.}\ }\textbf {\bibinfo {volume} {15}},\ \bibinfo
  {pages} {214} (\bibinfo {year} {2019})},\ \Eprint
  {https://arxiv.org/abs/1805.02973} {arXiv:1805.02973 [nucl-ex]} \BibitemShut
  {NoStop}%
\bibitem [{\citenamefont {Khachatryan}\ \emph {et~al.}(2016)\citenamefont
  {Khachatryan} \emph {et~al.}}]{CMS:2015fgy}%
  \BibitemOpen
  \bibfield  {author} {\bibinfo {author} {\bibfnamefont {V.}~\bibnamefont
  {Khachatryan}} \emph {et~al.} (\bibinfo {collaboration} {CMS}),\ }\href
  {https://doi.org/10.1103/PhysRevLett.116.172302} {\bibfield  {journal}
  {\bibinfo  {journal} {Phys. Rev. Lett.}\ }\textbf {\bibinfo {volume} {116}},\
  \bibinfo {pages} {172302} (\bibinfo {year} {2016})},\ \Eprint
  {https://arxiv.org/abs/1510.03068} {arXiv:1510.03068 [nucl-ex]} \BibitemShut
  {NoStop}%
\bibitem [{\citenamefont {Aad}\ \emph {et~al.}(2016)\citenamefont {Aad} \emph
  {et~al.}}]{ATLAS:2015hzw}%
  \BibitemOpen
  \bibfield  {author} {\bibinfo {author} {\bibfnamefont {G.}~\bibnamefont
  {Aad}} \emph {et~al.} (\bibinfo {collaboration} {ATLAS}),\ }\href
  {https://doi.org/10.1103/PhysRevLett.116.172301} {\bibfield  {journal}
  {\bibinfo  {journal} {Phys. Rev. Lett.}\ }\textbf {\bibinfo {volume} {116}},\
  \bibinfo {pages} {172301} (\bibinfo {year} {2016})},\ \Eprint
  {https://arxiv.org/abs/1509.04776} {arXiv:1509.04776 [hep-ex]} \BibitemShut
  {NoStop}%
\bibitem [{\citenamefont {Moreland}\ \emph {et~al.}(2020)\citenamefont
  {Moreland}, \citenamefont {Bernhard},\ and\ \citenamefont
  {Bass}}]{Moreland:2018gsh}%
  \BibitemOpen
  \bibfield  {author} {\bibinfo {author} {\bibfnamefont {J.~S.}\ \bibnamefont
  {Moreland}}, \bibinfo {author} {\bibfnamefont {J.~E.}\ \bibnamefont
  {Bernhard}},\ and\ \bibinfo {author} {\bibfnamefont {S.~A.}\ \bibnamefont
  {Bass}},\ }\href {https://doi.org/10.1103/PhysRevC.101.024911} {\bibfield
  {journal} {\bibinfo  {journal} {Phys. Rev. C}\ }\textbf {\bibinfo {volume}
  {101}},\ \bibinfo {pages} {024911} (\bibinfo {year} {2020})},\ \Eprint
  {https://arxiv.org/abs/1808.02106} {arXiv:1808.02106 [nucl-th]} \BibitemShut
  {NoStop}%
\bibitem [{\citenamefont {Bernhard}\ \emph {et~al.}(2019)\citenamefont
  {Bernhard}, \citenamefont {Moreland},\ and\ \citenamefont
  {Bass}}]{Bernhard:2019bmu}%
  \BibitemOpen
  \bibfield  {author} {\bibinfo {author} {\bibfnamefont {J.~E.}\ \bibnamefont
  {Bernhard}}, \bibinfo {author} {\bibfnamefont {J.~S.}\ \bibnamefont
  {Moreland}},\ and\ \bibinfo {author} {\bibfnamefont {S.~A.}\ \bibnamefont
  {Bass}},\ }\href {https://doi.org/10.1038/s41567-019-0611-8} {\bibfield
  {journal} {\bibinfo  {journal} {Nature Phys.}\ }\textbf {\bibinfo {volume}
  {15}},\ \bibinfo {pages} {1113} (\bibinfo {year} {2019})}\BibitemShut
  {NoStop}%
\bibitem [{\citenamefont {Everett}\ \emph
  {et~al.}(2021{\natexlab{a}})\citenamefont {Everett} \emph
  {et~al.}}]{JETSCAPE:2020shq}%
  \BibitemOpen
  \bibfield  {author} {\bibinfo {author} {\bibfnamefont {D.}~\bibnamefont
  {Everett}} \emph {et~al.} (\bibinfo {collaboration} {JETSCAPE}),\ }\href
  {https://doi.org/10.1103/PhysRevLett.126.242301} {\bibfield  {journal}
  {\bibinfo  {journal} {Phys. Rev. Lett.}\ }\textbf {\bibinfo {volume} {126}},\
  \bibinfo {pages} {242301} (\bibinfo {year} {2021}{\natexlab{a}})},\ \Eprint
  {https://arxiv.org/abs/2010.03928} {arXiv:2010.03928 [hep-ph]} \BibitemShut
  {NoStop}%
\bibitem [{\citenamefont {Everett}\ \emph
  {et~al.}(2021{\natexlab{b}})\citenamefont {Everett} \emph
  {et~al.}}]{JETSCAPE:2020mzn}%
  \BibitemOpen
  \bibfield  {author} {\bibinfo {author} {\bibfnamefont {D.}~\bibnamefont
  {Everett}} \emph {et~al.} (\bibinfo {collaboration} {JETSCAPE}),\ }\href
  {https://doi.org/10.1103/PhysRevC.103.054904} {\bibfield  {journal} {\bibinfo
   {journal} {Phys. Rev. C}\ }\textbf {\bibinfo {volume} {103}},\ \bibinfo
  {pages} {054904} (\bibinfo {year} {2021}{\natexlab{b}})},\ \Eprint
  {https://arxiv.org/abs/2011.01430} {arXiv:2011.01430 [hep-ph]} \BibitemShut
  {NoStop}%
\bibitem [{\citenamefont {Nijs}\ \emph
  {et~al.}(2021{\natexlab{a}})\citenamefont {Nijs}, \citenamefont {van~der
  Schee}, \citenamefont {G\"ursoy},\ and\ \citenamefont
  {Snellings}}]{Nijs:2020ors}%
  \BibitemOpen
  \bibfield  {author} {\bibinfo {author} {\bibfnamefont {G.}~\bibnamefont
  {Nijs}}, \bibinfo {author} {\bibfnamefont {W.}~\bibnamefont {van~der Schee}},
  \bibinfo {author} {\bibfnamefont {U.}~\bibnamefont {G\"ursoy}},\ and\
  \bibinfo {author} {\bibfnamefont {R.}~\bibnamefont {Snellings}},\ }\href
  {https://doi.org/10.1103/PhysRevLett.126.202301} {\bibfield  {journal}
  {\bibinfo  {journal} {Phys. Rev. Lett.}\ }\textbf {\bibinfo {volume} {126}},\
  \bibinfo {pages} {202301} (\bibinfo {year} {2021}{\natexlab{a}})},\ \Eprint
  {https://arxiv.org/abs/2010.15130} {arXiv:2010.15130 [nucl-th]} \BibitemShut
  {NoStop}%
\bibitem [{\citenamefont {Nijs}\ \emph
  {et~al.}(2021{\natexlab{b}})\citenamefont {Nijs}, \citenamefont {van~der
  Schee}, \citenamefont {G\"ursoy},\ and\ \citenamefont
  {Snellings}}]{Nijs:2020roc}%
  \BibitemOpen
  \bibfield  {author} {\bibinfo {author} {\bibfnamefont {G.}~\bibnamefont
  {Nijs}}, \bibinfo {author} {\bibfnamefont {W.}~\bibnamefont {van~der Schee}},
  \bibinfo {author} {\bibfnamefont {U.}~\bibnamefont {G\"ursoy}},\ and\
  \bibinfo {author} {\bibfnamefont {R.}~\bibnamefont {Snellings}},\ }\href
  {https://doi.org/10.1103/PhysRevC.103.054909} {\bibfield  {journal} {\bibinfo
   {journal} {Phys. Rev. C}\ }\textbf {\bibinfo {volume} {103}},\ \bibinfo
  {pages} {054909} (\bibinfo {year} {2021}{\natexlab{b}})},\ \Eprint
  {https://arxiv.org/abs/2010.15134} {arXiv:2010.15134 [nucl-th]} \BibitemShut
  {NoStop}%
\bibitem [{\citenamefont {Nijs}\ and\ \citenamefont {van~der
  Schee}(2022)}]{Nijs:2021clz}%
  \BibitemOpen
  \bibfield  {author} {\bibinfo {author} {\bibfnamefont {G.}~\bibnamefont
  {Nijs}}\ and\ \bibinfo {author} {\bibfnamefont {W.}~\bibnamefont {van~der
  Schee}},\ }\href {https://doi.org/10.1103/PhysRevC.106.044903} {\bibfield
  {journal} {\bibinfo  {journal} {Phys. Rev. C}\ }\textbf {\bibinfo {volume}
  {106}},\ \bibinfo {pages} {044903} (\bibinfo {year} {2022})},\ \Eprint
  {https://arxiv.org/abs/2110.13153} {arXiv:2110.13153 [nucl-th]} \BibitemShut
  {NoStop}%
\bibitem [{\citenamefont {Nijs}\ and\ \citenamefont {van~der
  Schee}()}]{Nijs:2023yab}%
  \BibitemOpen
  \bibfield  {author} {\bibinfo {author} {\bibfnamefont {G.}~\bibnamefont
  {Nijs}}\ and\ \bibinfo {author} {\bibfnamefont {W.}~\bibnamefont {van~der
  Schee}},\ }\href@noop {} {\ }\Eprint {https://arxiv.org/abs/2304.06191}
  {arXiv:2304.06191 [nucl-th]} \BibitemShut {NoStop}%
\bibitem [{\citenamefont {Heffernan}\ \emph {et~al.}()\citenamefont
  {Heffernan}, \citenamefont {Gale}, \citenamefont {Jeon},\ and\ \citenamefont
  {Paquet}}]{Heffernan:2023utr}%
  \BibitemOpen
  \bibfield  {author} {\bibinfo {author} {\bibfnamefont {M.~R.}\ \bibnamefont
  {Heffernan}}, \bibinfo {author} {\bibfnamefont {C.}~\bibnamefont {Gale}},
  \bibinfo {author} {\bibfnamefont {S.}~\bibnamefont {Jeon}},\ and\ \bibinfo
  {author} {\bibfnamefont {J.-F.}\ \bibnamefont {Paquet}},\ }\href@noop {} {\
  }\Eprint {https://arxiv.org/abs/2302.09478} {arXiv:2302.09478 [nucl-th]}
  \BibitemShut {NoStop}%
\bibitem [{\citenamefont {Liyanage}\ \emph {et~al.}()\citenamefont {Liyanage},
  \citenamefont {S\"urer}, \citenamefont {Plumlee}, \citenamefont {Wild},\ and\
  \citenamefont {Heinz}}]{Liyanage:2023nds}%
  \BibitemOpen
  \bibfield  {author} {\bibinfo {author} {\bibfnamefont {D.}~\bibnamefont
  {Liyanage}}, \bibinfo {author} {\bibfnamefont {O.}~\bibnamefont {S\"urer}},
  \bibinfo {author} {\bibfnamefont {M.}~\bibnamefont {Plumlee}}, \bibinfo
  {author} {\bibfnamefont {S.~M.}\ \bibnamefont {Wild}},\ and\ \bibinfo
  {author} {\bibfnamefont {U.}~\bibnamefont {Heinz}},\ }\href@noop {} {\
  }\Eprint {https://arxiv.org/abs/2302.14184} {arXiv:2302.14184 [nucl-th]}
  \BibitemShut {NoStop}%
\bibitem [{\citenamefont {Parkkila}\ \emph {et~al.}(2021)\citenamefont
  {Parkkila}, \citenamefont {Onnerstad},\ and\ \citenamefont
  {Kim}}]{Parkkila:2021tqq}%
  \BibitemOpen
  \bibfield  {author} {\bibinfo {author} {\bibfnamefont {J.~E.}\ \bibnamefont
  {Parkkila}}, \bibinfo {author} {\bibfnamefont {A.}~\bibnamefont
  {Onnerstad}},\ and\ \bibinfo {author} {\bibfnamefont {D.~J.}\ \bibnamefont
  {Kim}},\ }\href {https://doi.org/10.1103/PhysRevC.104.054904} {\bibfield
  {journal} {\bibinfo  {journal} {Phys. Rev. C}\ }\textbf {\bibinfo {volume}
  {104}},\ \bibinfo {pages} {054904} (\bibinfo {year} {2021})},\ \Eprint
  {https://arxiv.org/abs/2106.05019} {arXiv:2106.05019 [hep-ph]} \BibitemShut
  {NoStop}%
\bibitem [{\citenamefont {Parkkila}\ \emph {et~al.}(2022)\citenamefont
  {Parkkila}, \citenamefont {Onnerstad}, \citenamefont {Taghavi}, \citenamefont
  {Mordasini}, \citenamefont {Bilandzic}, \citenamefont {Virta},\ and\
  \citenamefont {Kim}}]{Parkkila:2021yha}%
  \BibitemOpen
  \bibfield  {author} {\bibinfo {author} {\bibfnamefont {J.~E.}\ \bibnamefont
  {Parkkila}}, \bibinfo {author} {\bibfnamefont {A.}~\bibnamefont {Onnerstad}},
  \bibinfo {author} {\bibfnamefont {S.~F.}\ \bibnamefont {Taghavi}}, \bibinfo
  {author} {\bibfnamefont {C.}~\bibnamefont {Mordasini}}, \bibinfo {author}
  {\bibfnamefont {A.}~\bibnamefont {Bilandzic}}, \bibinfo {author}
  {\bibfnamefont {M.}~\bibnamefont {Virta}},\ and\ \bibinfo {author}
  {\bibfnamefont {D.~J.}\ \bibnamefont {Kim}},\ }\href
  {https://doi.org/10.1016/j.physletb.2022.137485} {\bibfield  {journal}
  {\bibinfo  {journal} {Phys. Lett. B}\ }\textbf {\bibinfo {volume} {835}},\
  \bibinfo {pages} {137485} (\bibinfo {year} {2022})},\ \Eprint
  {https://arxiv.org/abs/2111.08145} {arXiv:2111.08145 [hep-ph]} \BibitemShut
  {NoStop}%
\bibitem [{\citenamefont {Pang}\ \emph {et~al.}(2018)\citenamefont {Pang},
  \citenamefont {Zhou}, \citenamefont {Su}, \citenamefont {Petersen},
  \citenamefont {St\"ocker},\ and\ \citenamefont {Wang}}]{Pang:2016vdc}%
  \BibitemOpen
  \bibfield  {author} {\bibinfo {author} {\bibfnamefont {L.-G.}\ \bibnamefont
  {Pang}}, \bibinfo {author} {\bibfnamefont {K.}~\bibnamefont {Zhou}}, \bibinfo
  {author} {\bibfnamefont {N.}~\bibnamefont {Su}}, \bibinfo {author}
  {\bibfnamefont {H.}~\bibnamefont {Petersen}}, \bibinfo {author}
  {\bibfnamefont {H.}~\bibnamefont {St\"ocker}},\ and\ \bibinfo {author}
  {\bibfnamefont {X.-N.}\ \bibnamefont {Wang}},\ }\href
  {https://doi.org/10.1038/s41467-017-02726-3} {\bibfield  {journal} {\bibinfo
  {journal} {Nature Commun.}\ }\textbf {\bibinfo {volume} {9}},\ \bibinfo
  {pages} {210} (\bibinfo {year} {2018})},\ \Eprint
  {https://arxiv.org/abs/1612.04262} {arXiv:1612.04262 [hep-ph]} \BibitemShut
  {NoStop}%
\bibitem [{\citenamefont {Steinheimer}\ \emph {et~al.}(2019)\citenamefont
  {Steinheimer}, \citenamefont {Pang}, \citenamefont {Zhou}, \citenamefont
  {Koch}, \citenamefont {Randrup},\ and\ \citenamefont
  {Stoecker}}]{Steinheimer:2019iso}%
  \BibitemOpen
  \bibfield  {author} {\bibinfo {author} {\bibfnamefont {J.}~\bibnamefont
  {Steinheimer}}, \bibinfo {author} {\bibfnamefont {L.}~\bibnamefont {Pang}},
  \bibinfo {author} {\bibfnamefont {K.}~\bibnamefont {Zhou}}, \bibinfo {author}
  {\bibfnamefont {V.}~\bibnamefont {Koch}}, \bibinfo {author} {\bibfnamefont
  {J.}~\bibnamefont {Randrup}},\ and\ \bibinfo {author} {\bibfnamefont
  {H.}~\bibnamefont {Stoecker}},\ }\href
  {https://doi.org/10.1007/JHEP12(2019)122} {\bibfield  {journal} {\bibinfo
  {journal} {J. High Energ. Phys}\ }\textbf {\bibinfo {volume} {12}},\ \bibinfo
  {pages} {122} (\bibinfo {year} {2019})},\ \Eprint
  {https://arxiv.org/abs/1906.06562} {arXiv:1906.06562 [nucl-th]} \BibitemShut
  {NoStop}%
\bibitem [{\citenamefont {Devetak}\ \emph {et~al.}(2020)\citenamefont
  {Devetak}, \citenamefont {Dubla}, \citenamefont {Floerchinger}, \citenamefont
  {Grossi}, \citenamefont {Masciocchi}, \citenamefont {Mazeliauskas},\ and\
  \citenamefont {Selyuzhenkov}}]{Devetak:2019lsk}%
  \BibitemOpen
  \bibfield  {author} {\bibinfo {author} {\bibfnamefont {D.}~\bibnamefont
  {Devetak}}, \bibinfo {author} {\bibfnamefont {A.}~\bibnamefont {Dubla}},
  \bibinfo {author} {\bibfnamefont {S.}~\bibnamefont {Floerchinger}}, \bibinfo
  {author} {\bibfnamefont {E.}~\bibnamefont {Grossi}}, \bibinfo {author}
  {\bibfnamefont {S.}~\bibnamefont {Masciocchi}}, \bibinfo {author}
  {\bibfnamefont {A.}~\bibnamefont {Mazeliauskas}},\ and\ \bibinfo {author}
  {\bibfnamefont {I.}~\bibnamefont {Selyuzhenkov}},\ }\href
  {https://doi.org/10.1007/JHEP06(2020)044} {\bibfield  {journal} {\bibinfo
  {journal} {J. High Energ. Phys}\ }\textbf {\bibinfo {volume} {06}},\ \bibinfo
  {pages} {044} (\bibinfo {year} {2020})},\ \Eprint
  {https://arxiv.org/abs/1909.10485} {arXiv:1909.10485 [hep-ph]} \BibitemShut
  {NoStop}%
\bibitem [{\citenamefont {Christiansen}\ \emph {et~al.}(2015)\citenamefont
  {Christiansen}, \citenamefont {Haas}, \citenamefont {Pawlowski},\ and\
  \citenamefont {Strodthoff}}]{Christiansen:2014ypa}%
  \BibitemOpen
  \bibfield  {author} {\bibinfo {author} {\bibfnamefont {N.}~\bibnamefont
  {Christiansen}}, \bibinfo {author} {\bibfnamefont {M.}~\bibnamefont {Haas}},
  \bibinfo {author} {\bibfnamefont {J.~M.}\ \bibnamefont {Pawlowski}},\ and\
  \bibinfo {author} {\bibfnamefont {N.}~\bibnamefont {Strodthoff}},\ }\href
  {https://doi.org/10.1103/PhysRevLett.115.112002} {\bibfield  {journal}
  {\bibinfo  {journal} {Phys. Rev. Lett.}\ }\textbf {\bibinfo {volume} {115}},\
  \bibinfo {pages} {112002} (\bibinfo {year} {2015})},\ \Eprint
  {https://arxiv.org/abs/1411.7986} {arXiv:1411.7986 [hep-ph]} \BibitemShut
  {NoStop}%
\bibitem [{\citenamefont {Pawlowski}(2022)}]{Pawlowski_private}%
  \BibitemOpen
  \bibfield  {author} {\bibinfo {author} {\bibfnamefont {J.~M.}\ \bibnamefont
  {Pawlowski}},\ }\href@noop {} {}\bibinfo {howpublished} {Private
  communication} (\bibinfo {year} {2022})\BibitemShut {NoStop}%
\bibitem [{\citenamefont {Moreland}\ \emph {et~al.}(2015)\citenamefont
  {Moreland}, \citenamefont {Bernhard},\ and\ \citenamefont
  {Bass}}]{Moreland:2014oya}%
  \BibitemOpen
  \bibfield  {author} {\bibinfo {author} {\bibfnamefont {J.~S.}\ \bibnamefont
  {Moreland}}, \bibinfo {author} {\bibfnamefont {J.~E.}\ \bibnamefont
  {Bernhard}},\ and\ \bibinfo {author} {\bibfnamefont {S.~A.}\ \bibnamefont
  {Bass}},\ }\href {https://doi.org/10.1103/PhysRevC.92.011901} {\bibfield
  {journal} {\bibinfo  {journal} {Phys. Rev. C}\ }\textbf {\bibinfo {volume}
  {92}},\ \bibinfo {pages} {011901} (\bibinfo {year} {2015})},\ \Eprint
  {https://arxiv.org/abs/1412.4708} {arXiv:1412.4708 [nucl-th]} \BibitemShut
  {NoStop}%
\bibitem [{\citenamefont {Floerchinger}\ \emph {et~al.}(2019)\citenamefont
  {Floerchinger}, \citenamefont {Grossi},\ and\ \citenamefont
  {Lion}}]{Floerchinger:2018pje}%
  \BibitemOpen
  \bibfield  {author} {\bibinfo {author} {\bibfnamefont {S.}~\bibnamefont
  {Floerchinger}}, \bibinfo {author} {\bibfnamefont {E.}~\bibnamefont
  {Grossi}},\ and\ \bibinfo {author} {\bibfnamefont {J.}~\bibnamefont {Lion}},\
  }\href {https://doi.org/10.1103/PhysRevC.100.014905} {\bibfield  {journal}
  {\bibinfo  {journal} {Phys. Rev. C}\ }\textbf {\bibinfo {volume} {100}},\
  \bibinfo {pages} {014905} (\bibinfo {year} {2019})},\ \Eprint
  {https://arxiv.org/abs/1811.01870} {arXiv:1811.01870 [nucl-th]} \BibitemShut
  {NoStop}%
\bibitem [{\citenamefont {Mazeliauskas}\ \emph {et~al.}(2019)\citenamefont
  {Mazeliauskas}, \citenamefont {Floerchinger}, \citenamefont {Grossi},\ and\
  \citenamefont {Teaney}}]{Mazeliauskas:2018irt}%
  \BibitemOpen
  \bibfield  {author} {\bibinfo {author} {\bibfnamefont {A.}~\bibnamefont
  {Mazeliauskas}}, \bibinfo {author} {\bibfnamefont {S.}~\bibnamefont
  {Floerchinger}}, \bibinfo {author} {\bibfnamefont {E.}~\bibnamefont
  {Grossi}},\ and\ \bibinfo {author} {\bibfnamefont {D.}~\bibnamefont
  {Teaney}},\ }\href {https://doi.org/10.1140/epjc/s10052-019-6791-7}
  {\bibfield  {journal} {\bibinfo  {journal} {Eur. Phys. J. C}\ }\textbf
  {\bibinfo {volume} {79}},\ \bibinfo {pages} {284} (\bibinfo {year} {2019})},\
  \Eprint {https://arxiv.org/abs/1809.11049} {arXiv:1809.11049 [nucl-th]}
  \BibitemShut {NoStop}%
\bibitem [{\citenamefont {Neal}(1996)}]{Neal1996}%
  \BibitemOpen
  \bibfield  {author} {\bibinfo {author} {\bibfnamefont {R.~M.}\ \bibnamefont
  {Neal}},\ }\bibinfo {title} {Priors for infinite networks},\ in\ \href
  {https://doi.org/10.1007/978-1-4612-0745-0_2} {\emph {\bibinfo {booktitle}
  {Bayesian Learning for Neural Networks}}}\ (\bibinfo  {publisher} {Springer
  New York},\ \bibinfo {year} {1996})\ pp.\ \bibinfo {pages}
  {29--53}\BibitemShut {NoStop}%
\bibitem [{\citenamefont {Capellino}\ \emph {et~al.}(2023)\citenamefont
  {Capellino}, \citenamefont {Dubla}, \citenamefont {Floerchinger},
  \citenamefont {Grossi}, \citenamefont {Kirchner},\ and\ \citenamefont
  {Masciocchi}}]{Capellino:2023cxe}%
  \BibitemOpen
  \bibfield  {author} {\bibinfo {author} {\bibfnamefont {F.}~\bibnamefont
  {Capellino}}, \bibinfo {author} {\bibfnamefont {A.}~\bibnamefont {Dubla}},
  \bibinfo {author} {\bibfnamefont {S.}~\bibnamefont {Floerchinger}}, \bibinfo
  {author} {\bibfnamefont {E.}~\bibnamefont {Grossi}}, \bibinfo {author}
  {\bibfnamefont {A.}~\bibnamefont {Kirchner}},\ and\ \bibinfo {author}
  {\bibfnamefont {S.}~\bibnamefont {Masciocchi}},\ }\href
  {https://doi.org/10.1103/PhysRevD.108.116011} {\bibfield  {journal} {\bibinfo
   {journal} {Phys. Rev. D}\ }\textbf {\bibinfo {volume} {108}},\ \bibinfo
  {pages} {116011} (\bibinfo {year} {2023})},\ \Eprint
  {https://arxiv.org/abs/2307.14449} {arXiv:2307.14449 [hep-ph]} \BibitemShut
  {NoStop}%
\bibitem [{\citenamefont {Capellino}\ \emph {et~al.}(2022)\citenamefont
  {Capellino}, \citenamefont {Beraudo}, \citenamefont {Dubla}, \citenamefont
  {Floerchinger}, \citenamefont {Masciocchi}, \citenamefont {Pawlowski},\ and\
  \citenamefont {Selyuzhenkov}}]{Capellino:2022nvf}%
  \BibitemOpen
  \bibfield  {author} {\bibinfo {author} {\bibfnamefont {F.}~\bibnamefont
  {Capellino}}, \bibinfo {author} {\bibfnamefont {A.}~\bibnamefont {Beraudo}},
  \bibinfo {author} {\bibfnamefont {A.}~\bibnamefont {Dubla}}, \bibinfo
  {author} {\bibfnamefont {S.}~\bibnamefont {Floerchinger}}, \bibinfo {author}
  {\bibfnamefont {S.}~\bibnamefont {Masciocchi}}, \bibinfo {author}
  {\bibfnamefont {J.}~\bibnamefont {Pawlowski}},\ and\ \bibinfo {author}
  {\bibfnamefont {I.}~\bibnamefont {Selyuzhenkov}},\ }\href
  {https://doi.org/10.1103/PhysRevD.106.034021} {\bibfield  {journal} {\bibinfo
   {journal} {Phys. Rev. D}\ }\textbf {\bibinfo {volume} {106}},\ \bibinfo
  {pages} {034021} (\bibinfo {year} {2022})},\ \Eprint
  {https://arxiv.org/abs/2205.07692} {arXiv:2205.07692 [nucl-th]} \BibitemShut
  {NoStop}%
\bibitem [{\citenamefont {Abelev}\ \emph
  {et~al.}(2013{\natexlab{a}})\citenamefont {Abelev} \emph
  {et~al.}}]{ALICE:2013mez}%
  \BibitemOpen
  \bibfield  {author} {\bibinfo {author} {\bibfnamefont {B.}~\bibnamefont
  {Abelev}} \emph {et~al.} (\bibinfo {collaboration} {ALICE}),\ }\href
  {https://doi.org/10.1103/PhysRevC.88.044910} {\bibfield  {journal} {\bibinfo
  {journal} {Phys. Rev. C}\ }\textbf {\bibinfo {volume} {88}},\ \bibinfo
  {pages} {044910} (\bibinfo {year} {2013}{\natexlab{a}})},\ \Eprint
  {https://arxiv.org/abs/1303.0737} {arXiv:1303.0737 [hep-ex]} \BibitemShut
  {NoStop}%
\bibitem [{\citenamefont {Abelev}\ \emph
  {et~al.}(2013{\natexlab{b}})\citenamefont {Abelev} \emph
  {et~al.}}]{ALICE:2013cdo}%
  \BibitemOpen
  \bibfield  {author} {\bibinfo {author} {\bibfnamefont {B.~B.}\ \bibnamefont
  {Abelev}} \emph {et~al.} (\bibinfo {collaboration} {ALICE}),\ }\href
  {https://doi.org/10.1103/PhysRevLett.111.222301} {\bibfield  {journal}
  {\bibinfo  {journal} {Phys. Rev. Lett.}\ }\textbf {\bibinfo {volume} {111}},\
  \bibinfo {pages} {222301} (\bibinfo {year} {2013}{\natexlab{b}})},\ \Eprint
  {https://arxiv.org/abs/1307.5530} {arXiv:1307.5530 [nucl-ex]} \BibitemShut
  {NoStop}%
\bibitem [{\citenamefont {Acharya}\ \emph {et~al.}(2020)\citenamefont {Acharya}
  \emph {et~al.}}]{ALICE:2019hno}%
  \BibitemOpen
  \bibfield  {author} {\bibinfo {author} {\bibfnamefont {S.}~\bibnamefont
  {Acharya}} \emph {et~al.} (\bibinfo {collaboration} {ALICE}),\ }\href
  {https://doi.org/10.1103/PhysRevC.101.044907} {\bibfield  {journal} {\bibinfo
   {journal} {Phys. Rev. C}\ }\textbf {\bibinfo {volume} {101}},\ \bibinfo
  {pages} {044907} (\bibinfo {year} {2020})},\ \Eprint
  {https://arxiv.org/abs/1910.07678} {arXiv:1910.07678 [nucl-ex]} \BibitemShut
  {NoStop}%
\bibitem [{\citenamefont {Acharya}\ \emph {et~al.}(2021)\citenamefont {Acharya}
  \emph {et~al.}}]{ALICE:2021lsv}%
  \BibitemOpen
  \bibfield  {author} {\bibinfo {author} {\bibfnamefont {S.}~\bibnamefont
  {Acharya}} \emph {et~al.} (\bibinfo {collaboration} {ALICE}),\ }\href
  {https://doi.org/10.1140/epjc/s10052-021-09304-4} {\bibfield  {journal}
  {\bibinfo  {journal} {Eur. Phys. J. C}\ }\textbf {\bibinfo {volume} {81}},\
  \bibinfo {pages} {584} (\bibinfo {year} {2021})},\ \Eprint
  {https://arxiv.org/abs/2101.03100} {arXiv:2101.03100 [nucl-ex]} \BibitemShut
  {NoStop}%
\bibitem [{\citenamefont {Giacalone}()}]{Giacalone:2022hnz}%
  \BibitemOpen
  \bibfield  {author} {\bibinfo {author} {\bibfnamefont {G.}~\bibnamefont
  {Giacalone}},\ }\href@noop {} {\ }\Eprint {https://arxiv.org/abs/2208.06839}
  {arXiv:2208.06839 [nucl-th]} \BibitemShut {NoStop}%
\bibitem [{\citenamefont {Acharya}\ \emph {et~al.}(2018)\citenamefont {Acharya}
  \emph {et~al.}}]{ALICE-PUBLIC-2018-011}%
  \BibitemOpen
  \bibfield  {author} {\bibinfo {author} {\bibfnamefont {S.}~\bibnamefont
  {Acharya}} \emph {et~al.} (\bibinfo {collaboration} {ALICE}),\ }\href
  {https://cds.cern.ch/record/2636623} {\bibfield  {journal} {\bibinfo
  {journal} {ALICE-PUBLIC-2018-011}\ } (\bibinfo {year} {2018})}\BibitemShut
  {NoStop}%
\bibitem [{\citenamefont {Bally}\ \emph {et~al.}(2022)\citenamefont {Bally},
  \citenamefont {Bender}, \citenamefont {Giacalone},\ and\ \citenamefont
  {Som\`a}}]{Bally:2021qys}%
  \BibitemOpen
  \bibfield  {author} {\bibinfo {author} {\bibfnamefont {B.}~\bibnamefont
  {Bally}}, \bibinfo {author} {\bibfnamefont {M.}~\bibnamefont {Bender}},
  \bibinfo {author} {\bibfnamefont {G.}~\bibnamefont {Giacalone}},\ and\
  \bibinfo {author} {\bibfnamefont {V.}~\bibnamefont {Som\`a}},\ }\href
  {https://doi.org/10.1103/PhysRevLett.128.082301} {\bibfield  {journal}
  {\bibinfo  {journal} {Phys. Rev. Lett.}\ }\textbf {\bibinfo {volume} {128}},\
  \bibinfo {pages} {082301} (\bibinfo {year} {2022})},\ \Eprint
  {https://arxiv.org/abs/2108.09578} {arXiv:2108.09578 [nucl-th]} \BibitemShut
  {NoStop}%
\bibitem [{\citenamefont {Giacalone}(2020)}]{Giacalone:2020ymy}%
  \BibitemOpen
  \bibfield  {author} {\bibinfo {author} {\bibfnamefont {G.}~\bibnamefont
  {Giacalone}},\ }\emph {\bibinfo {title} {{A matter of shape: seeing the
  deformation of atomic nuclei at high-energy colliders}}},\ \href@noop {}
  {Ph.D. thesis},\ \bibinfo  {school} {U. Paris-Saclay} (\bibinfo {year}
  {2020}),\ \Eprint {https://arxiv.org/abs/2101.00168} {arXiv:2101.00168
  [nucl-th]} \BibitemShut {NoStop}%
\bibitem [{\citenamefont {Floerchinger}\ and\ \citenamefont
  {Wiedemann}(2014{\natexlab{a}})}]{Floerchinger:2013rya}%
  \BibitemOpen
  \bibfield  {author} {\bibinfo {author} {\bibfnamefont {S.}~\bibnamefont
  {Floerchinger}}\ and\ \bibinfo {author} {\bibfnamefont {U.~A.}\ \bibnamefont
  {Wiedemann}},\ }\href {https://doi.org/10.1016/j.physletb.2013.12.025}
  {\bibfield  {journal} {\bibinfo  {journal} {Phys. Lett. B}\ }\textbf
  {\bibinfo {volume} {728}},\ \bibinfo {pages} {407} (\bibinfo {year}
  {2014}{\natexlab{a}})},\ \Eprint {https://arxiv.org/abs/1307.3453}
  {arXiv:1307.3453 [hep-ph]} \BibitemShut {NoStop}%
\bibitem [{\citenamefont {Floerchinger}\ and\ \citenamefont
  {Wiedemann}(2014{\natexlab{b}})}]{Floerchinger:2013hza}%
  \BibitemOpen
  \bibfield  {author} {\bibinfo {author} {\bibfnamefont {S.}~\bibnamefont
  {Floerchinger}}\ and\ \bibinfo {author} {\bibfnamefont {U.~A.}\ \bibnamefont
  {Wiedemann}},\ }\href {https://doi.org/10.1103/PhysRevC.89.034914} {\bibfield
   {journal} {\bibinfo  {journal} {Phys. Rev. C}\ }\textbf {\bibinfo {volume}
  {89}},\ \bibinfo {pages} {034914} (\bibinfo {year} {2014}{\natexlab{b}})},\
  \Eprint {https://arxiv.org/abs/1311.7613} {arXiv:1311.7613 [hep-ph]}
  \BibitemShut {NoStop}%
\bibitem [{\citenamefont {Floerchinger}\ and\ \citenamefont
  {Wiedemann}(2014{\natexlab{c}})}]{Floerchinger:2014fta}%
  \BibitemOpen
  \bibfield  {author} {\bibinfo {author} {\bibfnamefont {S.}~\bibnamefont
  {Floerchinger}}\ and\ \bibinfo {author} {\bibfnamefont {U.~A.}\ \bibnamefont
  {Wiedemann}},\ }\href {https://doi.org/10.1007/JHEP08(2014)005} {\bibfield
  {journal} {\bibinfo  {journal} {J. High Energ. Phys}\ }\textbf {\bibinfo
  {volume} {08}},\ \bibinfo {pages} {005} (\bibinfo {year}
  {2014}{\natexlab{c}})},\ \Eprint {https://arxiv.org/abs/1405.4393}
  {arXiv:1405.4393 [hep-ph]} \BibitemShut {NoStop}%
\bibitem [{\citenamefont {Floerchinger}\ and\ \citenamefont
  {Grossi}(2018)}]{Floerchinger:2017cii}%
  \BibitemOpen
  \bibfield  {author} {\bibinfo {author} {\bibfnamefont {S.}~\bibnamefont
  {Floerchinger}}\ and\ \bibinfo {author} {\bibfnamefont {E.}~\bibnamefont
  {Grossi}},\ }\href {https://doi.org/10.1007/JHEP08(2018)186} {\bibfield
  {journal} {\bibinfo  {journal} {J. High Energ. Phys}\ }\textbf {\bibinfo
  {volume} {08}},\ \bibinfo {pages} {186} (\bibinfo {year} {2018})},\ \Eprint
  {https://arxiv.org/abs/1711.06687} {arXiv:1711.06687 [nucl-th]} \BibitemShut
  {NoStop}%
\bibitem [{\citenamefont {Demir}\ and\ \citenamefont
  {Bass}(2009)}]{Demir:2008tr}%
  \BibitemOpen
  \bibfield  {author} {\bibinfo {author} {\bibfnamefont {N.}~\bibnamefont
  {Demir}}\ and\ \bibinfo {author} {\bibfnamefont {S.~A.}\ \bibnamefont
  {Bass}},\ }\href {https://doi.org/10.1103/PhysRevLett.102.172302} {\bibfield
  {journal} {\bibinfo  {journal} {Phys. Rev. Lett.}\ }\textbf {\bibinfo
  {volume} {102}},\ \bibinfo {pages} {172302} (\bibinfo {year} {2009})},\
  \Eprint {https://arxiv.org/abs/0812.2422} {arXiv:0812.2422 [nucl-th]}
  \BibitemShut {NoStop}%
\bibitem [{\citenamefont {Rose}\ \emph {et~al.}(2018)\citenamefont {Rose},
  \citenamefont {Torres-Rincon}, \citenamefont {Sch\"afer}, \citenamefont
  {Oliinychenko},\ and\ \citenamefont {Petersen}}]{Rose:2017bjz}%
  \BibitemOpen
  \bibfield  {author} {\bibinfo {author} {\bibfnamefont {J.~B.}\ \bibnamefont
  {Rose}}, \bibinfo {author} {\bibfnamefont {J.~M.}\ \bibnamefont
  {Torres-Rincon}}, \bibinfo {author} {\bibfnamefont {A.}~\bibnamefont
  {Sch\"afer}}, \bibinfo {author} {\bibfnamefont {D.~R.}\ \bibnamefont
  {Oliinychenko}},\ and\ \bibinfo {author} {\bibfnamefont {H.}~\bibnamefont
  {Petersen}},\ }\href {https://doi.org/10.1103/PhysRevC.97.055204} {\bibfield
  {journal} {\bibinfo  {journal} {Phys. Rev. C}\ }\textbf {\bibinfo {volume}
  {97}},\ \bibinfo {pages} {055204} (\bibinfo {year} {2018})},\ \Eprint
  {https://arxiv.org/abs/1709.03826} {arXiv:1709.03826 [nucl-th]} \BibitemShut
  {NoStop}%
\bibitem [{\citenamefont {Karsch}\ \emph {et~al.}(2008)\citenamefont {Karsch},
  \citenamefont {Kharzeev},\ and\ \citenamefont {Tuchin}}]{Karsch:2007jc}%
  \BibitemOpen
  \bibfield  {author} {\bibinfo {author} {\bibfnamefont {F.}~\bibnamefont
  {Karsch}}, \bibinfo {author} {\bibfnamefont {D.}~\bibnamefont {Kharzeev}},\
  and\ \bibinfo {author} {\bibfnamefont {K.}~\bibnamefont {Tuchin}},\ }\href
  {https://doi.org/10.1016/j.physletb.2008.01.080} {\bibfield  {journal}
  {\bibinfo  {journal} {Phys. Lett. B}\ }\textbf {\bibinfo {volume} {663}},\
  \bibinfo {pages} {217} (\bibinfo {year} {2008})},\ \Eprint
  {https://arxiv.org/abs/0711.0914} {arXiv:0711.0914 [hep-ph]} \BibitemShut
  {NoStop}%
\bibitem [{\citenamefont {Noronha-Hostler}\ \emph {et~al.}(2009)\citenamefont
  {Noronha-Hostler}, \citenamefont {Noronha},\ and\ \citenamefont
  {Greiner}}]{Noronha-Hostler:2008kkf}%
  \BibitemOpen
  \bibfield  {author} {\bibinfo {author} {\bibfnamefont {J.}~\bibnamefont
  {Noronha-Hostler}}, \bibinfo {author} {\bibfnamefont {J.}~\bibnamefont
  {Noronha}},\ and\ \bibinfo {author} {\bibfnamefont {C.}~\bibnamefont
  {Greiner}},\ }\href {https://doi.org/10.1103/PhysRevLett.103.172302}
  {\bibfield  {journal} {\bibinfo  {journal} {Phys. Rev. Lett.}\ }\textbf
  {\bibinfo {volume} {103}},\ \bibinfo {pages} {172302} (\bibinfo {year}
  {2009})},\ \Eprint {https://arxiv.org/abs/0811.1571} {arXiv:0811.1571
  [nucl-th]} \BibitemShut {NoStop}%
\bibitem [{\citenamefont {Denicol}\ \emph {et~al.}(2009)\citenamefont
  {Denicol}, \citenamefont {Kodama}, \citenamefont {Koide},\ and\ \citenamefont
  {Mota}}]{Denicol:2009am}%
  \BibitemOpen
  \bibfield  {author} {\bibinfo {author} {\bibfnamefont {G.~S.}\ \bibnamefont
  {Denicol}}, \bibinfo {author} {\bibfnamefont {T.}~\bibnamefont {Kodama}},
  \bibinfo {author} {\bibfnamefont {T.}~\bibnamefont {Koide}},\ and\ \bibinfo
  {author} {\bibfnamefont {P.}~\bibnamefont {Mota}},\ }\href
  {https://doi.org/10.1103/PhysRevC.80.064901} {\bibfield  {journal} {\bibinfo
  {journal} {Phys. Rev. C}\ }\textbf {\bibinfo {volume} {80}},\ \bibinfo
  {pages} {064901} (\bibinfo {year} {2009})},\ \Eprint
  {https://arxiv.org/abs/0903.3595} {arXiv:0903.3595 [hep-ph]} \BibitemShut
  {NoStop}%
\bibitem [{\citenamefont {Astrakhantsev}\ \emph {et~al.}(2018)\citenamefont
  {Astrakhantsev}, \citenamefont {Braguta},\ and\ \citenamefont
  {Kotov}}]{Astrakhantsev:2018oue}%
  \BibitemOpen
  \bibfield  {author} {\bibinfo {author} {\bibfnamefont {N.~Y.}\ \bibnamefont
  {Astrakhantsev}}, \bibinfo {author} {\bibfnamefont {V.~V.}\ \bibnamefont
  {Braguta}},\ and\ \bibinfo {author} {\bibfnamefont {A.~Y.}\ \bibnamefont
  {Kotov}},\ }\href {https://doi.org/10.1103/PhysRevD.98.054515} {\bibfield
  {journal} {\bibinfo  {journal} {Phys. Rev. D}\ }\textbf {\bibinfo {volume}
  {98}},\ \bibinfo {pages} {054515} (\bibinfo {year} {2018})},\ \Eprint
  {https://arxiv.org/abs/1804.02382} {arXiv:1804.02382 [hep-lat]} \BibitemShut
  {NoStop}%
\bibitem [{\citenamefont {Lu}\ and\ \citenamefont {Moore}(2011)}]{Lu:2011df}%
  \BibitemOpen
  \bibfield  {author} {\bibinfo {author} {\bibfnamefont {E.}~\bibnamefont
  {Lu}}\ and\ \bibinfo {author} {\bibfnamefont {G.~D.}\ \bibnamefont {Moore}},\
  }\href {https://doi.org/10.1103/PhysRevC.83.044901} {\bibfield  {journal}
  {\bibinfo  {journal} {Phys. Rev. C}\ }\textbf {\bibinfo {volume} {83}},\
  \bibinfo {pages} {044901} (\bibinfo {year} {2011})},\ \Eprint
  {https://arxiv.org/abs/1102.0017} {arXiv:1102.0017 [hep-ph]} \BibitemShut
  {NoStop}%
\bibitem [{\citenamefont {Teaney}(2003)}]{Teaney:2003kp}%
  \BibitemOpen
  \bibfield  {author} {\bibinfo {author} {\bibfnamefont {D.}~\bibnamefont
  {Teaney}},\ }\href {https://doi.org/10.1103/PhysRevC.68.034913} {\bibfield
  {journal} {\bibinfo  {journal} {Phys. Rev.}\ }\textbf {\bibinfo {volume}
  {C68}},\ \bibinfo {pages} {034913} (\bibinfo {year} {2003})},\ \Eprint
  {https://arxiv.org/abs/nucl-th/0301099} {arXiv:nucl-th/0301099 [nucl-th]}
  \BibitemShut {NoStop}%
\bibitem [{\citenamefont {Paquet}\ \emph {et~al.}(2016)\citenamefont {Paquet},
  \citenamefont {Shen}, \citenamefont {Denicol}, \citenamefont {Luzum},
  \citenamefont {Schenke}, \citenamefont {Jeon},\ and\ \citenamefont
  {Gale}}]{Paquet:2015lta}%
  \BibitemOpen
  \bibfield  {author} {\bibinfo {author} {\bibfnamefont {J.-F.}\ \bibnamefont
  {Paquet}}, \bibinfo {author} {\bibfnamefont {C.}~\bibnamefont {Shen}},
  \bibinfo {author} {\bibfnamefont {G.~S.}\ \bibnamefont {Denicol}}, \bibinfo
  {author} {\bibfnamefont {M.}~\bibnamefont {Luzum}}, \bibinfo {author}
  {\bibfnamefont {B.}~\bibnamefont {Schenke}}, \bibinfo {author} {\bibfnamefont
  {S.}~\bibnamefont {Jeon}},\ and\ \bibinfo {author} {\bibfnamefont
  {C.}~\bibnamefont {Gale}},\ }\href
  {https://doi.org/10.1103/PhysRevC.93.044906} {\bibfield  {journal} {\bibinfo
  {journal} {Phys. Rev.}\ }\textbf {\bibinfo {volume} {C93}},\ \bibinfo {pages}
  {044906} (\bibinfo {year} {2016})},\ \Eprint
  {https://arxiv.org/abs/1509.06738} {arXiv:1509.06738 [hep-ph]} \BibitemShut
  {NoStop}%
\bibitem [{\citenamefont {Denicol}\ \emph {et~al.}(2014)\citenamefont
  {Denicol}, \citenamefont {Jeon},\ and\ \citenamefont
  {Gale}}]{Denicol:2014vaa}%
  \BibitemOpen
  \bibfield  {author} {\bibinfo {author} {\bibfnamefont {G.~S.}\ \bibnamefont
  {Denicol}}, \bibinfo {author} {\bibfnamefont {S.}~\bibnamefont {Jeon}},\ and\
  \bibinfo {author} {\bibfnamefont {C.}~\bibnamefont {Gale}},\ }\href
  {https://doi.org/10.1103/PhysRevC.90.024912} {\bibfield  {journal} {\bibinfo
  {journal} {Phys. Rev. C}\ }\textbf {\bibinfo {volume} {90}},\ \bibinfo
  {pages} {024912} (\bibinfo {year} {2014})},\ \Eprint
  {https://arxiv.org/abs/1403.0962} {arXiv:1403.0962 [nucl-th]} \BibitemShut
  {NoStop}%
\bibitem [{\citenamefont {Cooper}\ and\ \citenamefont
  {Frye}(1974)}]{Cooper:1974mv}%
  \BibitemOpen
  \bibfield  {author} {\bibinfo {author} {\bibfnamefont {F.}~\bibnamefont
  {Cooper}}\ and\ \bibinfo {author} {\bibfnamefont {G.}~\bibnamefont {Frye}},\
  }\href {https://doi.org/10.1103/PhysRevD.10.186} {\bibfield  {journal}
  {\bibinfo  {journal} {Phys. Rev. D}\ }\textbf {\bibinfo {volume} {10}},\
  \bibinfo {pages} {186} (\bibinfo {year} {1974})}\BibitemShut {NoStop}%
\bibitem [{\citenamefont {Bebie}\ \emph {et~al.}(1992)\citenamefont {Bebie},
  \citenamefont {Gerber}, \citenamefont {Goity},\ and\ \citenamefont
  {Leutwyler}}]{Bebie:1991ij}%
  \BibitemOpen
  \bibfield  {author} {\bibinfo {author} {\bibfnamefont {H.}~\bibnamefont
  {Bebie}}, \bibinfo {author} {\bibfnamefont {P.}~\bibnamefont {Gerber}},
  \bibinfo {author} {\bibfnamefont {J.~L.}\ \bibnamefont {Goity}},\ and\
  \bibinfo {author} {\bibfnamefont {H.}~\bibnamefont {Leutwyler}},\ }\href
  {https://doi.org/10.1016/0550-3213(92)90005-V} {\bibfield  {journal}
  {\bibinfo  {journal} {Nucl. Phys.}\ }\textbf {\bibinfo {volume} {B378}},\
  \bibinfo {pages} {95} (\bibinfo {year} {1992})}\BibitemShut {NoStop}%
\bibitem [{\citenamefont {Huovinen}(2008)}]{Huovinen:2007xh}%
  \BibitemOpen
  \bibfield  {author} {\bibinfo {author} {\bibfnamefont {P.}~\bibnamefont
  {Huovinen}},\ }\href {https://doi.org/10.1140/epja/i2007-10611-3} {\bibfield
  {journal} {\bibinfo  {journal} {Eur. Phys. J. A}\ }\textbf {\bibinfo {volume}
  {37}},\ \bibinfo {pages} {121} (\bibinfo {year} {2008})},\ \Eprint
  {https://arxiv.org/abs/0710.4379} {arXiv:0710.4379 [nucl-th]} \BibitemShut
  {NoStop}%
\bibitem [{\citenamefont {Aamodt}\ \emph {et~al.}(2011)\citenamefont {Aamodt}
  \emph {et~al.}}]{ALICE:2011dyt}%
  \BibitemOpen
  \bibfield  {author} {\bibinfo {author} {\bibfnamefont {K.}~\bibnamefont
  {Aamodt}} \emph {et~al.} (\bibinfo {collaboration} {ALICE}),\ }\href
  {https://doi.org/10.1016/j.physletb.2010.12.053} {\bibfield  {journal}
  {\bibinfo  {journal} {Phys. Lett. B}\ }\textbf {\bibinfo {volume} {696}},\
  \bibinfo {pages} {328} (\bibinfo {year} {2011})},\ \Eprint
  {https://arxiv.org/abs/1012.4035} {arXiv:1012.4035 [nucl-ex]} \BibitemShut
  {NoStop}%
\bibitem [{\citenamefont {Alba}\ \emph {et~al.}(2017)\citenamefont {Alba} \emph
  {et~al.}}]{Alba:2017mqu}%
  \BibitemOpen
  \bibfield  {author} {\bibinfo {author} {\bibfnamefont {P.}~\bibnamefont
  {Alba}} \emph {et~al.},\ }\href {https://doi.org/10.1103/PhysRevD.96.034517}
  {\bibfield  {journal} {\bibinfo  {journal} {Phys. Rev.}\ }\textbf {\bibinfo
  {volume} {D96}},\ \bibinfo {pages} {034517} (\bibinfo {year} {2017})},\
  \Eprint {https://arxiv.org/abs/1702.01113} {arXiv:1702.01113 [hep-lat]}
  \BibitemShut {NoStop}%
\bibitem [{\citenamefont {Alba}\ \emph {et~al.}(2018)\citenamefont {Alba},
  \citenamefont {Mantovani~Sarti}, \citenamefont {Noronha}, \citenamefont
  {Noronha-Hostler}, \citenamefont {Parotto}, \citenamefont
  {Portillo~Vazquez},\ and\ \citenamefont {Ratti}}]{Alba:2017hhe}%
  \BibitemOpen
  \bibfield  {author} {\bibinfo {author} {\bibfnamefont {P.}~\bibnamefont
  {Alba}}, \bibinfo {author} {\bibfnamefont {V.}~\bibnamefont
  {Mantovani~Sarti}}, \bibinfo {author} {\bibfnamefont {J.}~\bibnamefont
  {Noronha}}, \bibinfo {author} {\bibfnamefont {J.}~\bibnamefont
  {Noronha-Hostler}}, \bibinfo {author} {\bibfnamefont {P.}~\bibnamefont
  {Parotto}}, \bibinfo {author} {\bibfnamefont {I.}~\bibnamefont
  {Portillo~Vazquez}},\ and\ \bibinfo {author} {\bibfnamefont {C.}~\bibnamefont
  {Ratti}},\ }\href {https://doi.org/10.1103/PhysRevC.98.034909} {\bibfield
  {journal} {\bibinfo  {journal} {Phys. Rev.}\ }\textbf {\bibinfo {volume}
  {C98}},\ \bibinfo {pages} {034909} (\bibinfo {year} {2018})},\ \Eprint
  {https://arxiv.org/abs/1711.05207} {arXiv:1711.05207 [nucl-th]} \BibitemShut
  {NoStop}%
\bibitem [{\citenamefont {Parotto}(2019)}]{Parotto_private}%
  \BibitemOpen
  \bibfield  {author} {\bibinfo {author} {\bibfnamefont {P.}~\bibnamefont
  {Parotto}},\ }\href@noop {} {}\bibinfo {howpublished} {Private communication}
  (\bibinfo {year} {2019})\BibitemShut {NoStop}%
\bibitem [{\citenamefont {Ryu}\ \emph {et~al.}(2018)\citenamefont {Ryu} \emph
  {et~al.}}]{Ryu:2017qzn}%
  \BibitemOpen
  \bibfield  {author} {\bibinfo {author} {\bibfnamefont {S.}~\bibnamefont
  {Ryu}} \emph {et~al.},\ }\href {https://doi.org/10.1103/PhysRevC.97.034910}
  {\bibfield  {journal} {\bibinfo  {journal} {Phys. Rev. C}\ }\textbf {\bibinfo
  {volume} {97}},\ \bibinfo {pages} {034910} (\bibinfo {year} {2018})},\
  \Eprint {https://arxiv.org/abs/1704.04216} {arXiv:1704.04216 [nucl-th]}
  \BibitemShut {NoStop}%
\bibitem [{\citenamefont {Andronic}\ \emph {et~al.}(2018)\citenamefont
  {Andronic}, \citenamefont {Braun-Munzinger}, \citenamefont {Redlich},\ and\
  \citenamefont {Stachel}}]{Andronic:2017pug}%
  \BibitemOpen
  \bibfield  {author} {\bibinfo {author} {\bibfnamefont {A.}~\bibnamefont
  {Andronic}}, \bibinfo {author} {\bibfnamefont {P.}~\bibnamefont
  {Braun-Munzinger}}, \bibinfo {author} {\bibfnamefont {K.}~\bibnamefont
  {Redlich}},\ and\ \bibinfo {author} {\bibfnamefont {J.}~\bibnamefont
  {Stachel}},\ }\href {https://doi.org/10.1038/s41586-018-0491-6} {\bibfield
  {journal} {\bibinfo  {journal} {Nature}\ }\textbf {\bibinfo {volume} {561}},\
  \bibinfo {pages} {321} (\bibinfo {year} {2018})},\ \Eprint
  {https://arxiv.org/abs/1710.09425} {arXiv:1710.09425 [nucl-th]} \BibitemShut
  {NoStop}%
\bibitem [{\citenamefont {Seemann}(2022)}]{ThesisYannis}%
  \BibitemOpen
  \bibfield  {author} {\bibinfo {author} {\bibfnamefont {Y.}~\bibnamefont
  {Seemann}},\ }\emph {\bibinfo {title} {Multidimensional parameter
  optimization in fluid dynamic simulations of heavy-ion collisions}},\ \href
  {https://www.physi.uni-heidelberg.de/Publications/MasterThesisYannisSeemann.pdf}
  {Master's thesis},\ \bibinfo  {school} {Universit{\"a}t Heidelberg} (\bibinfo
  {year} {2022})\BibitemShut {NoStop}%
\bibitem [{\citenamefont {Paszke}\ \emph {et~al.}(2019)\citenamefont {Paszke}
  \emph {et~al.}}]{paszke2019pytorch}%
  \BibitemOpen
  \bibfield  {author} {\bibinfo {author} {\bibfnamefont {A.}~\bibnamefont
  {Paszke}} \emph {et~al.},\ }\href@noop {} {\bibfield  {journal} {\bibinfo
  {journal} {Advances in neural information processing systems}\ }\textbf
  {\bibinfo {volume} {32}} (\bibinfo {year} {2019})}\BibitemShut {NoStop}%
\bibitem [{\citenamefont {Liaw}\ \emph {et~al.}()\citenamefont {Liaw},
  \citenamefont {Liang}, \citenamefont {Nishihara}, \citenamefont {Moritz},
  \citenamefont {Gonzalez},\ and\ \citenamefont {Stoica}}]{liaw2018tune}%
  \BibitemOpen
  \bibfield  {author} {\bibinfo {author} {\bibfnamefont {R.}~\bibnamefont
  {Liaw}}, \bibinfo {author} {\bibfnamefont {E.}~\bibnamefont {Liang}},
  \bibinfo {author} {\bibfnamefont {R.}~\bibnamefont {Nishihara}}, \bibinfo
  {author} {\bibfnamefont {P.}~\bibnamefont {Moritz}}, \bibinfo {author}
  {\bibfnamefont {J.~E.}\ \bibnamefont {Gonzalez}},\ and\ \bibinfo {author}
  {\bibfnamefont {I.}~\bibnamefont {Stoica}},\ }\href@noop {} {\ }\Eprint
  {https://arxiv.org/abs/1807.05118} {arXiv:1807.05118 [cs.LG]} \BibitemShut
  {NoStop}%
\bibitem [{\citenamefont {Foreman-Mackey}\ \emph {et~al.}(2013)\citenamefont
  {Foreman-Mackey}, \citenamefont {Hogg}, \citenamefont {Lang},\ and\
  \citenamefont {Goodman}}]{Foreman-Mackey:2012any}%
  \BibitemOpen
  \bibfield  {author} {\bibinfo {author} {\bibfnamefont {D.}~\bibnamefont
  {Foreman-Mackey}}, \bibinfo {author} {\bibfnamefont {D.~W.}\ \bibnamefont
  {Hogg}}, \bibinfo {author} {\bibfnamefont {D.}~\bibnamefont {Lang}},\ and\
  \bibinfo {author} {\bibfnamefont {J.}~\bibnamefont {Goodman}},\ }\href
  {https://doi.org/10.1086/670067} {\bibfield  {journal} {\bibinfo  {journal}
  {Publ. Astron. Soc. Pac.}\ }\textbf {\bibinfo {volume} {125}},\ \bibinfo
  {pages} {306} (\bibinfo {year} {2013})},\ \Eprint
  {https://arxiv.org/abs/1202.3665} {arXiv:1202.3665 [astro-ph.IM]}
  \BibitemShut {NoStop}%
\bibitem [{\citenamefont {Gale}\ \emph
  {et~al.}(2013{\natexlab{a}})\citenamefont {Gale}, \citenamefont {Jeon},\ and\
  \citenamefont {Schenke}}]{Gale:2013da}%
  \BibitemOpen
  \bibfield  {author} {\bibinfo {author} {\bibfnamefont {C.}~\bibnamefont
  {Gale}}, \bibinfo {author} {\bibfnamefont {S.}~\bibnamefont {Jeon}},\ and\
  \bibinfo {author} {\bibfnamefont {B.}~\bibnamefont {Schenke}},\ }\href
  {https://doi.org/10.1142/S0217751X13400113} {\bibfield  {journal} {\bibinfo
  {journal} {Int. J. Mod. Phys. A}\ }\textbf {\bibinfo {volume} {28}},\
  \bibinfo {pages} {1340011} (\bibinfo {year} {2013}{\natexlab{a}})},\ \Eprint
  {https://arxiv.org/abs/1301.5893} {arXiv:1301.5893 [nucl-th]} \BibitemShut
  {NoStop}%
\bibitem [{\citenamefont {Bernhard}\ \emph {et~al.}(2016)\citenamefont
  {Bernhard}, \citenamefont {Moreland}, \citenamefont {Bass}, \citenamefont
  {Liu},\ and\ \citenamefont {Heinz}}]{Bernhard:2016tnd}%
  \BibitemOpen
  \bibfield  {author} {\bibinfo {author} {\bibfnamefont {J.~E.}\ \bibnamefont
  {Bernhard}}, \bibinfo {author} {\bibfnamefont {J.~S.}\ \bibnamefont
  {Moreland}}, \bibinfo {author} {\bibfnamefont {S.~A.}\ \bibnamefont {Bass}},
  \bibinfo {author} {\bibfnamefont {J.}~\bibnamefont {Liu}},\ and\ \bibinfo
  {author} {\bibfnamefont {U.}~\bibnamefont {Heinz}},\ }\href
  {https://doi.org/10.1103/PhysRevC.94.024907} {\bibfield  {journal} {\bibinfo
  {journal} {Phys. Rev. C}\ }\textbf {\bibinfo {volume} {94}},\ \bibinfo
  {pages} {024907} (\bibinfo {year} {2016})},\ \Eprint
  {https://arxiv.org/abs/1605.03954} {arXiv:1605.03954 [nucl-th]} \BibitemShut
  {NoStop}%
\bibitem [{\citenamefont {Teaney}(2010)}]{Teaney:2009qa}%
  \BibitemOpen
  \bibfield  {author} {\bibinfo {author} {\bibfnamefont {D.~A.}\ \bibnamefont
  {Teaney}},\ }\bibinfo {title} {{Viscous Hydrodynamics and the Quark Gluon
  Plasma}},\ in\ \href {https://doi.org/10.1142/9789814293297_0004} {\emph
  {\bibinfo {booktitle} {{Quark-gluon plasma 4}}}}\ (\bibinfo  {publisher}
  {World Scientific},\ \bibinfo {year} {2010})\ pp.\ \bibinfo {pages}
  {207--266},\ \Eprint {https://arxiv.org/abs/0905.2433} {arXiv:0905.2433
  [nucl-th]} \BibitemShut {NoStop}%
\bibitem [{\citenamefont {Dubla}\ \emph {et~al.}(2018)\citenamefont {Dubla},
  \citenamefont {Masciocchi}, \citenamefont {Pawlowski}, \citenamefont
  {Schenke}, \citenamefont {Shen},\ and\ \citenamefont
  {Stachel}}]{Dubla:2018czx}%
  \BibitemOpen
  \bibfield  {author} {\bibinfo {author} {\bibfnamefont {A.}~\bibnamefont
  {Dubla}}, \bibinfo {author} {\bibfnamefont {S.}~\bibnamefont {Masciocchi}},
  \bibinfo {author} {\bibfnamefont {J.~M.}\ \bibnamefont {Pawlowski}}, \bibinfo
  {author} {\bibfnamefont {B.}~\bibnamefont {Schenke}}, \bibinfo {author}
  {\bibfnamefont {C.}~\bibnamefont {Shen}},\ and\ \bibinfo {author}
  {\bibfnamefont {J.}~\bibnamefont {Stachel}},\ }\href
  {https://doi.org/10.1016/j.nuclphysa.2018.09.046} {\bibfield  {journal}
  {\bibinfo  {journal} {Nucl. Phys. A}\ }\textbf {\bibinfo {volume} {979}},\
  \bibinfo {pages} {251} (\bibinfo {year} {2018})},\ \Eprint
  {https://arxiv.org/abs/1805.02985} {arXiv:1805.02985 [nucl-th]} \BibitemShut
  {NoStop}%
\bibitem [{\citenamefont {Gale}\ \emph
  {et~al.}(2013{\natexlab{b}})\citenamefont {Gale}, \citenamefont {Jeon},
  \citenamefont {Schenke}, \citenamefont {Tribedy},\ and\ \citenamefont
  {Venugopalan}}]{Gale:2012rq}%
  \BibitemOpen
  \bibfield  {author} {\bibinfo {author} {\bibfnamefont {C.}~\bibnamefont
  {Gale}}, \bibinfo {author} {\bibfnamefont {S.}~\bibnamefont {Jeon}}, \bibinfo
  {author} {\bibfnamefont {B.}~\bibnamefont {Schenke}}, \bibinfo {author}
  {\bibfnamefont {P.}~\bibnamefont {Tribedy}},\ and\ \bibinfo {author}
  {\bibfnamefont {R.}~\bibnamefont {Venugopalan}},\ }\href
  {https://doi.org/10.1103/PhysRevLett.110.012302} {\bibfield  {journal}
  {\bibinfo  {journal} {Phys. Rev. Lett.}\ }\textbf {\bibinfo {volume} {110}},\
  \bibinfo {pages} {012302} (\bibinfo {year} {2013}{\natexlab{b}})},\ \Eprint
  {https://arxiv.org/abs/1209.6330} {arXiv:1209.6330 [nucl-th]} \BibitemShut
  {NoStop}%
\bibitem [{\citenamefont {Begun}\ and\ \citenamefont
  {Florkowski}(2015)}]{Begun:2015ifa}%
  \BibitemOpen
  \bibfield  {author} {\bibinfo {author} {\bibfnamefont {V.}~\bibnamefont
  {Begun}}\ and\ \bibinfo {author} {\bibfnamefont {W.}~\bibnamefont
  {Florkowski}},\ }\href {https://doi.org/10.1103/PhysRevC.91.054909}
  {\bibfield  {journal} {\bibinfo  {journal} {Phys. Rev. C}\ }\textbf {\bibinfo
  {volume} {91}},\ \bibinfo {pages} {054909} (\bibinfo {year} {2015})},\
  \Eprint {https://arxiv.org/abs/1503.04040} {arXiv:1503.04040 [nucl-th]}
  \BibitemShut {NoStop}%
\bibitem [{\citenamefont {Begun}(2016)}]{Begun:2016cva}%
  \BibitemOpen
  \bibfield  {author} {\bibinfo {author} {\bibfnamefont {V.}~\bibnamefont
  {Begun}},\ }\href {https://doi.org/10.1103/PhysRevC.94.054904} {\bibfield
  {journal} {\bibinfo  {journal} {Phys. Rev. C}\ }\textbf {\bibinfo {volume}
  {94}},\ \bibinfo {pages} {054904} (\bibinfo {year} {2016})},\ \Eprint
  {https://arxiv.org/abs/1603.02254} {arXiv:1603.02254 [nucl-th]} \BibitemShut
  {NoStop}%
\bibitem [{\citenamefont {Schnedermann}\ \emph {et~al.}(1993)\citenamefont
  {Schnedermann}, \citenamefont {Sollfrank},\ and\ \citenamefont
  {Heinz}}]{Schnedermann:1993ws}%
  \BibitemOpen
  \bibfield  {author} {\bibinfo {author} {\bibfnamefont {E.}~\bibnamefont
  {Schnedermann}}, \bibinfo {author} {\bibfnamefont {J.}~\bibnamefont
  {Sollfrank}},\ and\ \bibinfo {author} {\bibfnamefont {U.~W.}\ \bibnamefont
  {Heinz}},\ }\href {https://doi.org/10.1103/PhysRevC.48.2462} {\bibfield
  {journal} {\bibinfo  {journal} {Phys. Rev. C}\ }\textbf {\bibinfo {volume}
  {48}},\ \bibinfo {pages} {2462} (\bibinfo {year} {1993})},\ \Eprint
  {https://arxiv.org/abs/nucl-th/9307020} {arXiv:nucl-th/9307020} \BibitemShut
  {NoStop}%
\bibitem [{\citenamefont {Huovinen}\ \emph {et~al.}(2017)\citenamefont
  {Huovinen}, \citenamefont {Lo}, \citenamefont {Marczenko}, \citenamefont
  {Morita}, \citenamefont {Redlich},\ and\ \citenamefont
  {Sasaki}}]{Huovinen:2016xxq}%
  \BibitemOpen
  \bibfield  {author} {\bibinfo {author} {\bibfnamefont {P.}~\bibnamefont
  {Huovinen}}, \bibinfo {author} {\bibfnamefont {P.~M.}\ \bibnamefont {Lo}},
  \bibinfo {author} {\bibfnamefont {M.}~\bibnamefont {Marczenko}}, \bibinfo
  {author} {\bibfnamefont {K.}~\bibnamefont {Morita}}, \bibinfo {author}
  {\bibfnamefont {K.}~\bibnamefont {Redlich}},\ and\ \bibinfo {author}
  {\bibfnamefont {C.}~\bibnamefont {Sasaki}},\ }\href
  {https://doi.org/10.1016/j.physletb.2017.03.060} {\bibfield  {journal}
  {\bibinfo  {journal} {Phys. Lett. B}\ }\textbf {\bibinfo {volume} {769}},\
  \bibinfo {pages} {509} (\bibinfo {year} {2017})},\ \Eprint
  {https://arxiv.org/abs/1608.06817} {arXiv:1608.06817 [hep-ph]} \BibitemShut
  {NoStop}%
\bibitem [{\citenamefont {Grossi}\ \emph {et~al.}(2021)\citenamefont {Grossi},
  \citenamefont {Soloviev}, \citenamefont {Teaney},\ and\ \citenamefont
  {Yan}}]{Grossi:2021gqi}%
  \BibitemOpen
  \bibfield  {author} {\bibinfo {author} {\bibfnamefont {E.}~\bibnamefont
  {Grossi}}, \bibinfo {author} {\bibfnamefont {A.}~\bibnamefont {Soloviev}},
  \bibinfo {author} {\bibfnamefont {D.}~\bibnamefont {Teaney}},\ and\ \bibinfo
  {author} {\bibfnamefont {F.}~\bibnamefont {Yan}},\ }\href
  {https://doi.org/10.1103/PhysRevD.104.034025} {\bibfield  {journal} {\bibinfo
   {journal} {Phys. Rev. D}\ }\textbf {\bibinfo {volume} {104}},\ \bibinfo
  {pages} {034025} (\bibinfo {year} {2021})},\ \Eprint
  {https://arxiv.org/abs/2101.10847} {arXiv:2101.10847 [nucl-th]} \BibitemShut
  {NoStop}%
\bibitem [{\citenamefont {van Hecke}\ \emph {et~al.}(1999)\citenamefont {van
  Hecke}, \citenamefont {Sorge},\ and\ \citenamefont {Xu}}]{vanHecke:1999jh}%
  \BibitemOpen
  \bibfield  {author} {\bibinfo {author} {\bibfnamefont {H.}~\bibnamefont {van
  Hecke}}, \bibinfo {author} {\bibfnamefont {H.}~\bibnamefont {Sorge}},\ and\
  \bibinfo {author} {\bibfnamefont {N.}~\bibnamefont {Xu}},\ }\href
  {https://doi.org/10.1016/S0375-9474(99)85073-8} {\bibfield  {journal}
  {\bibinfo  {journal} {Nucl. Phys. A}\ }\textbf {\bibinfo {volume} {661}},\
  \bibinfo {pages} {493} (\bibinfo {year} {1999})}\BibitemShut {NoStop}%
\bibitem [{\citenamefont {Arbex}\ \emph {et~al.}(2001)\citenamefont {Arbex},
  \citenamefont {Grassi}, \citenamefont {Hama},\ and\ \citenamefont
  {Socolowski}}]{Arbex:2001vx}%
  \BibitemOpen
  \bibfield  {author} {\bibinfo {author} {\bibfnamefont {N.}~\bibnamefont
  {Arbex}}, \bibinfo {author} {\bibfnamefont {F.}~\bibnamefont {Grassi}},
  \bibinfo {author} {\bibfnamefont {Y.}~\bibnamefont {Hama}},\ and\ \bibinfo
  {author} {\bibfnamefont {O.}~\bibnamefont {Socolowski}},\ }\href
  {https://doi.org/10.1103/PhysRevC.64.064906} {\bibfield  {journal} {\bibinfo
  {journal} {Phys. Rev. C}\ }\textbf {\bibinfo {volume} {64}},\ \bibinfo
  {pages} {064906} (\bibinfo {year} {2001})}\BibitemShut {NoStop}%
\bibitem [{\citenamefont {Chatterjee}\ \emph {et~al.}(2013)\citenamefont
  {Chatterjee}, \citenamefont {Godbole},\ and\ \citenamefont
  {Gupta}}]{Chatterjee:2013yga}%
  \BibitemOpen
  \bibfield  {author} {\bibinfo {author} {\bibfnamefont {S.}~\bibnamefont
  {Chatterjee}}, \bibinfo {author} {\bibfnamefont {R.~M.}\ \bibnamefont
  {Godbole}},\ and\ \bibinfo {author} {\bibfnamefont {S.}~\bibnamefont
  {Gupta}},\ }\href {https://doi.org/10.1016/j.physletb.2013.11.008} {\bibfield
   {journal} {\bibinfo  {journal} {Phys. Lett. B}\ }\textbf {\bibinfo {volume}
  {727}},\ \bibinfo {pages} {554} (\bibinfo {year} {2013})},\ \Eprint
  {https://arxiv.org/abs/1306.2006} {arXiv:1306.2006 [nucl-th]} \BibitemShut
  {NoStop}%
\bibitem [{\citenamefont {Bellwied}\ \emph {et~al.}(2013)\citenamefont
  {Bellwied}, \citenamefont {Borsanyi}, \citenamefont {Fodor}, \citenamefont
  {Katz},\ and\ \citenamefont {Ratti}}]{Bellwied:2013cta}%
  \BibitemOpen
  \bibfield  {author} {\bibinfo {author} {\bibfnamefont {R.}~\bibnamefont
  {Bellwied}}, \bibinfo {author} {\bibfnamefont {S.}~\bibnamefont {Borsanyi}},
  \bibinfo {author} {\bibfnamefont {Z.}~\bibnamefont {Fodor}}, \bibinfo
  {author} {\bibfnamefont {S.~D.}\ \bibnamefont {Katz}},\ and\ \bibinfo
  {author} {\bibfnamefont {C.}~\bibnamefont {Ratti}},\ }\href
  {https://doi.org/10.1103/PhysRevLett.111.202302} {\bibfield  {journal}
  {\bibinfo  {journal} {Phys. Rev. Lett.}\ }\textbf {\bibinfo {volume} {111}},\
  \bibinfo {pages} {202302} (\bibinfo {year} {2013})},\ \Eprint
  {https://arxiv.org/abs/1305.6297} {arXiv:1305.6297 [hep-lat]} \BibitemShut
  {NoStop}%
\bibitem [{\citenamefont {Kirchner}\ \emph {et~al.}()\citenamefont {Kirchner},
  \citenamefont {Grossi},\ and\ \citenamefont
  {Floerchinger}}]{Kirchner:2023fsj}%
  \BibitemOpen
  \bibfield  {author} {\bibinfo {author} {\bibfnamefont {A.}~\bibnamefont
  {Kirchner}}, \bibinfo {author} {\bibfnamefont {E.}~\bibnamefont {Grossi}},\
  and\ \bibinfo {author} {\bibfnamefont {S.}~\bibnamefont {Floerchinger}},\
  }\href@noop {} {\ }\Eprint {https://arxiv.org/abs/2308.10616}
  {arXiv:2308.10616 [hep-ph]} \BibitemShut {NoStop}%
\bibitem [{\citenamefont {Song}\ \emph {et~al.}(2011)\citenamefont {Song},
  \citenamefont {Bass},\ and\ \citenamefont {Heinz}}]{Song:2010aq}%
  \BibitemOpen
  \bibfield  {author} {\bibinfo {author} {\bibfnamefont {H.}~\bibnamefont
  {Song}}, \bibinfo {author} {\bibfnamefont {S.~A.}\ \bibnamefont {Bass}},\
  and\ \bibinfo {author} {\bibfnamefont {U.}~\bibnamefont {Heinz}},\ }\href
  {https://doi.org/10.1103/PhysRevC.83.024912} {\bibfield  {journal} {\bibinfo
  {journal} {Phys. Rev. C}\ }\textbf {\bibinfo {volume} {83}},\ \bibinfo
  {pages} {024912} (\bibinfo {year} {2011})},\ \Eprint
  {https://arxiv.org/abs/1012.0555} {arXiv:1012.0555 [nucl-th]} \BibitemShut
  {NoStop}%
\end{thebibliography}%

\end{document}